\documentclass[12pt]{article}
\usepackage{xr-hyper}
\makeatletter

\usepackage{longtable}
\usepackage{xcolor}
\usepackage{enumitem}
\usepackage{amssymb}
\usepackage[]{natbib}
\usepackage{amsmath,amsfonts,bm}
\usepackage[parfill]{parskip}
\usepackage{amsthm}
\usepackage[toc,page]{appendix}
\newcommand{\bM}{\boldsymbol{M}}
\newcommand{\1}{\boldsymbol{1}}

\usepackage[pagewise]{lineno}

\usepackage{ragged2e}

\usepackage{microtype}
\usepackage{graphicx}
\usepackage{booktabs} % for professional tables
\usepackage{multirow} 

\usepackage{comment}
\usepackage{caption}
\usepackage{bbm}
\usepackage{subcaption}
\usepackage{enumitem}
\setitemize{noitemsep,topsep=0pt,parsep=0pt,partopsep=0pt}

\usepackage{algorithm}
\usepackage{chngcntr} % for figure labels in appendix
\usepackage[noend]{algpseudocode}
\algrenewcommand\textproc{}% Used to be \textsc
\algdef{SE}[SUBALG]{Indent}{EndIndent}{}{\algorithmicend\ }%
\algtext*{Indent}
\algtext*{EndIndent}

\newcommand{\Mean}{{\mathbb{E}}}

\newcommand{\prob}{{\mathbb{P}}}

\newcommand{\btheta}{\boldsymbol{\theta}}
\newcommand{\bTheta}{\boldsymbol{\Theta}}

\newcommand{\blambda}{\boldsymbol{\lambda}}

\newcommand{\bR}{\boldsymbol{R}}

\newcommand{\bI}{\boldsymbol{I}}

\newcommand{\bB}{\boldsymbol{B}}

\newcommand{\bPhi}{\boldsymbol{\Phi}}

\newcommand{\bbeta}{\boldsymbol{\beta}}

\DeclareMathOperator*{\argmin}{arg\,min}
\DeclareMathOperator*{\argmax}{arg\,max}

\newcommand{\bgamma}{\boldsymbol{\gamma}}

\newtheorem{theorem}{Theorem}
\newtheorem{remark}{Remark}

\newtheorem{prop}{Proposition}

\newtheorem{asmp}{Assumption}

\usepackage{threeparttable}

\newcommand{\E}{\mathbb{E}}

\newcommand{\balpha}{\boldsymbol{\alpha}}
\newcommand{\bP}{\boldsymbol{P}}

\newcommand{\bS}{\boldsymbol{S}}
\numberwithin{equation}{section}
\usepackage{hyperref}

\hypersetup{
	unicode=false,
	pdftoolbar=true,
	pdfmenubar=true,
	pdffitwindow=false,     % window fit to page when opened
	pdfstartview={FitH},    % fits the width of the page to the window
	pdfkeywords={}, % list of keywords
	pdfnewwindow=true,      % links in new window
	colorlinks=true,       % false: boxed links; true: colored links
	linkcolor=cyan,          % color of internal links (change box color with linkbordercolor)
	citecolor=blue,        % color of links to bibliography
	filecolor=cyan,      % color of file links
	urlcolor=cyan           % color of external links
}

\usepackage[utf8]{inputenc}

\newcommand{\neighbor}[1]%
{\overline{#1}}

\newcommand{\blind}{1}
\usepackage{colortbl}
\definecolor{lightblue}{rgb}{0.8, 0.9, 1}
\usepackage{mathtools}
\graphicspath{{fig/}}
\DeclarePairedDelimiter\norm{\lVert}{\rVert}

\begin{document}

\def\spacingset#1{\renewcommand{\baselinestretch}%
{#1}\small\normalsize} \spacingset{1.4}

\if1\blind
{
  \title{ \bf Off-policy Evaluation in Doubly Inhomogeneous Environments}
  \author{Zeyu Bian$^{1}$, 
Chengchun Shi$^2$, Zhengling Qi$^{3}$ and Lan Wang$^{1}$ \\ 
\smallskip\\
$^1$Department of Management Science, University of Miami\\ $^2$Department of Statistics, London School of Economics and Political Science \\ $^3$ Department of Decision Sciences, The George Washington University}
\date{}
  \maketitle
} \fi

\if0\blind
{
  \bigskip
  \bigskip
  \bigskip
  \begin{center}
    {\LARGE\bf Off-policy Evaluation in Doubly Inhomogeneous Environments}
\end{center}
  \medskip
} \fi
	
\bigskip

\begin{abstract}
This work aims to study off-policy evaluation (OPE) under scenarios where two key reinforcement learning (RL) assumptions --  temporal stationarity and individual homogeneity are both violated. To handle the ``double inhomogeneities", we propose a class of latent factor models for the reward and transition functions, under which we develop a general OPE framework that consists of both model-based and model-free approaches. To our knowledge, this is the first paper that develops statistically sound OPE methods in offline RL with double inhomogeneities. It contributes to a deeper understanding of OPE in environments, where standard RL assumptions are not met, and provides several practical approaches in these settings. We establish the theoretical properties of the proposed value estimators and empirically show that our approach outperforms state-of-the-art methods. %that ignore either temporal nonstationarity or individual heterogeneity. 
Finally, we illustrate our method on a data set from the Medical Information Mart for Intensive Care. An R implementation of the proposed procedure is available at \url{https://github.com/ZeyuBian/2FEOPE}.
\end{abstract}

\noindent%
\vfill

\newpage
\spacingset{1.7} 

\section{Introduction}

%In sequential decision making, 
Reinforcement learning \citep[RL,][]{sutton2018} aims to optimize an agent's long-term reward by learning an optimal policy that determines the best action to take under every circumstance. %, i.e., given a situation, learn the best action to maximize some pre-defined numerical reward. 
RL is closely related to the dynamic treatment regimens (DTR) or adaptive treatment strategies in statistical research for precision medicine \citep{Murphy,gest,qian2011performance,kosorok2015adaptive,shi2018high,tsiatis2019dynamic,qi2020multi,zhou2022estimating}, which seeks to obtain the optimal treatment policy in finite horizon settings with a few treatment stages that maximizes patients' expected outcome. Nevertheless, statistical methods for DTR mentioned above normally cannot handle large or infinite horizon settings. They require the number of trajectories to tend to infinity to achieve estimation consistency, unlike RL, which works even with finite number of trajectories under certain conditions. In addition to precision medicine, RL has been applied to various fields, such as games \citep{silver2016mastering},  ridesharing \citep{xu2018large}, mobile health \citep{liao2021off} and robotics \citep{levine2020offline}. 

In this article, we focus on off-policy evaluation (OPE), whose objective is to evaluate the value function of a given target policy using data collected from a potentially different policy, known as the behavior policy. OPE is important in applications in which directly implementing a policy involves potential risks and high costs. For instance, in healthcare, it would be expensive to conduct a randomized experiment to recruit many individuals and follow them up for the duration of the entire experiment. Meanwhile, it might be unethical to directly apply a new treatment policy to some individuals without offline validation. 
It is therefore important to develop RL methods only using historical data, and OPE is particularly vital in offline RL. Generally speaking, existing OPE methods can be divided into four categories: model-based methods \citep{gottesman2019combining,zhang2021autoregressive}, importance sampling methods \citep{precup2000eligibility,liu2018breaking,wang2021projected}, direct methods \citep{luckett2020estimating,liao2021off,shi2020statistical}, and doubly robust methods \citep{jiang2016doubly,uehara2020minimax,kallus2022efficiently,liao2022batch}. %which typically are a fusion of the former two approaches. 
See \cite{uehara2022review} and the references therein for an overview.

\textbf{\textit{Motivation}}. Most methods in the RL literature rely on the following two critical assumptions: temporal stationarity and individual homogeneity. The 
temporal stationarity assumption requires that the system dynamics %(i.e., observation transition and reward functions) 
for each subject do not depend on the time whereas individual homogeneity requires the system dynamics at each time to be identical across all subjects. Nonetheless, both conditions are likely to be violated in many RL applications, e.g., mobile health and infectious disease control \citep{hu2022doubly}. This work draws partial motivation from the longitudinal data of septic patients obtained from the Medical Information Mart for Intensive Care %(MIMIC-III) 
\citep[MIMIC-III,][]{johnson2016mimic}, a database containing information on critical care patients. Sepsis is a severe and potentially fatal condition that occurs when the human body's response to an infection injures its own tissues and organs \citep{singer2016third}. It can progress rapidly and cause multiple organ failures, resulting in %a decline in a patient's health and 
an increased risk of death. Prompt treatment of sepsis is thus essential for improving patient outcomes and reducing mortality rates. However, the heterogeneity in patients' response to sepsis treatments \citep{evans2021surviving}, as well as a potentially non-stationary environment (the data includes patients' medical information over 10 years) make it challenging to effectively manage the illness using existing RL methods. Our analysis provides insights into evaluating the impact of different treatment strategies, %on patient outcomes 
facilitating the development of effective and personalized approaches to sepsis care.

In the statistical literature, \citet{li2022reinforcement} and \citet{wang2023robust} developed  hypothesis testing procedures to assess the stationarity assumption in RL, based on which policy learning procedures were proposed to handle possibly nonstationary environments. \citet{chen2022reinforcement} developed a transferred Q-learning algorithm and an auto-clustered policy iteration algorithm to handle heterogeneous data. However, these methods require either temporal stationarity or individual homogeneity, and would fail in doubly inhomogeneous environments when both assumptions are violated. %mentioned above only considered the non-stationarity case in RL, leaving the scenario of subject heterogeneity unstudied. 
\citet{hu2022doubly} proposed an algorithm to adaptively split the data into rectangles in which the system dynamics are identical over time and across individuals. They studied policy learning instead of OPE. In addition, they %require local stationarity and homogeneity to 
imposed a latent group structure over time and population. 
This structural assumption can be violated when the dynamics vary smoothly over both population and time. %see Remark \ref{remark:hu} in supplementary article for details. 

\begin{comment}
Under a doubly inhomogeneous environment, it is also worth thinking about what long-term numerical quantity should be investigated. Most existing literature related to OPE focus on the discounted sum of rewards \citep{jiang2016doubly,liu2018breaking,kallus2020double,shi2020statistical,kallus2022efficiently} across the population, 
which require a discount factor to balance the trade-off between immediate and future rewards: as the discount factor becomes larger (approaches to 1), the agent puts more emphasis on future rewards. There are also other cases in which the authors consider evaluating the long-term average reward of a target policy \citep{liao2021off,shi2022multi,liao2022batch} across the population. They argue that the average reward might better represent the long-term performance of a target policy, especially in mobile health and ride sharing where a sufficiently long time horizon can be common. To account for the impact of subject heterogeneity and nonstationarity, we suggest evaluate the following four quantities (see formal definition in the later section): \begin{enumerate}
\item the average reward following the target policy $\eta^{\pi}$, where $\pi$ is the target policy;
    \item the $i$th subject's average reward aggregated over time following the target policy $\eta^{\pi}_{i}$;
    \item the average reward in the population at time $t$ following the target policy $\eta^{\pi}_{t}$;
    \item the $i$th subject's average reward at time $t$ following the target policy $\eta^{\pi}_{i,t}$.
\end{enumerate}
\end{comment}

\textbf{\textit{Challenges}}. 
OPE is substantially more challenging
under the doubly inhomogeneous environments. First, the evaluation target is different. In particular, most existing solutions developed in doubly homogeneous environments have predominantly focused on evaluating the expected long-term reward following the target policy aggregated over time and population. In contrast, the following four time- and/or individual-specific values are of particular interest in the presence of double inhomogeneities: \begin{enumerate}
    \item The expected long-term reward aggregated over both time and population;\vspace{-1em}
    \item The expected long-term reward aggregated over time for a given subject;
    \vspace{-1em}
    \item The expected reward at a given time aggregated over population;
    \vspace{-1em}
    \item The expected reward at a given time for a given subject.
\end{enumerate} Second, an unresolved challenge is
how to efficiently borrow information over time and population for OPE. On one hand, to account for the subject heterogeneity or temporal nonstationarity, %problem and to estimate $\eta^{\pi}_{i}$, 
one could conduct OPE based on the data within each individual trajectory or at a given time. However, this approach may result in an estimator with a high variance.
On the other hand, naively pulling data over population and time without careful considerations would lead to biased estimators. 
%existing advanced OPE methods based on its own $i$th data trajectory; nevertheless, this may yield a loss of statistical efficiency. Furthermore, when the number of time periods is small, the resulting estimator can be unstable and noisy. On the other hand, to deal with the nonstationarity case, the natural choice is to use the idea of backward induction \citep{bellman1957markovian,fudenberg1991game,bather2000decision}, i.e., starting at the end, iteratively evaluate the reward backwards in time. However, when the number of time points increases, it might suffer a large variance \citep{voloshin2019empirical}; it also remains challenging how to estimate the desired quantity iteratively under individual heterogeneity. To address these challenges, we propose a general framework under the doubly inhomogeneous scenario. Moreover, we studied both model-based and model-free approaches. Unlike the standard OPE methods that only use data from each time point or each trajectory, our proposed model-free estimator allows us to efficiently use pre-collected data from all trajectories across all time periods to estimate the reward function.

\textbf{\textit{Contributions}}. This work makes the following contributions. First, to our knowledge, it is the first study to investigate OPE in doubly inhomogeneous RL domains. Unlike prior works that primarily focused on evaluating the average effect over time and population, we provide a systematic approach for examining values that are specific to time and/or individuals. These values hold particular importance in the context of double inhomogeneities. 
%in contrast to previous works in the RL literature that only studied either non-stationarity setting or the individual heterogeneity case, our work is concerned with OPE with scenario in which both stationarity and individual homogeneity assumptions can be violated. In addition, unlike the existing literature, three different value functions mentioned above are investigated in this paper, hence it could provide a more comprehensive perspective and a more nuanced understanding on the topic of OPE. To the best of our knowledge, this is the first work to develop an OPE approach in the field of offline RL focus on the double inhomogeneous setting. Probably most relevant work to us is \citet{hu2022doubly}, where they discussed the topic of policy learning rather than the OPE under the double inhomogeneous environment. Moreover, \citet{hu2022doubly} presumed that the system dynamic is only temporal non-stationarity and further assumed there exists a latent group structure among individuals.
% in contrast, our proposed method does not impose any temporal and latent group structure type assumptions.

Second, we present a comprehensive framework for doubly inhomogeneous OPE which comprises both model-free and model-based methods. To effectively utilize information in the presence of temporal nonstationarity and individual heterogeneity, we introduce a class of two-way doubly inhomogeneous decision process (TWDIDP) models and develop corresponding policy value estimators. Our proposal shares similar spirits with the two-way fixed effects model that is widely studied in economics and social science \citep{angrist2009mostly,imai2021use}. Nonetheless, our model is substantially more complicated %in that 
due to the incorporation of carryover effects: in our model, the current treatment not only affects its immediate outcome, but also impacts the future outcomes through its effect on the future observation via the transition function. In contrast, the fixed effects models commonly employed in the panel data literature tend to exclude carryover effects \citep{imai2019should,arkhangelsky2021double}.

Finally, we systematically investigate the theoretical properties of the proposed model-free method. In particular, we derive the convergence rates of various proposed value estimators, showing that the estimated average effect, individual-specific effect, time-specific effect and individual- and time-specific effect converge at a rate of $(NT)^{-1/2}$, $T^{-1/2}$, $N^{-1/2}$ and $\min^{-1/2}(N,T)$, respectively, up to some logarithmic factors, where $N$ is the number of trajectories and $T$ is the number of time points. We further establish the limiting distributions of these estimators.

\textbf{\textit{Organization}}. The rest of this paper is organized as follows. In Section \ref{sec:pre}, we introduce the proposed doubly inhomogeneous decision process model to 
incorporate temporal nonstationarity and individual heterogeneity. 
In Sections \ref{sec:mdfree} and \ref{sec:mdbased}, we present our proposed model-free and model-based methods. We analyze their statistical properties in Section \ref{sec:theory}. A series of comprehensive simulation studies are conducted in Section \ref{sec:sims}. Finally, in Section \ref{sec:real}, we 
illustrate the proposed approach using the MIMIC-III dataset.

\section{Two-way Doubly Inhomogeneous Decision Processes}\label{sec:pre}
\textbf{\textit{Data}}. We first describe the dataset. We assume the offline data consists of $N$ independent trajectories, each with $T$ many time points, and can be summarized as the following observation-action-reward triplets $\{(O_{i,t}, A_{i,t}, R_{i,t}): 1\le i\le N, 1\le t\le T\}$ where $i$ indexes the $i$th individual and $t$ indexes the $t$th time point. For example, in mobile health applications, $O_{i,t} \in \mathbb{R}^{d}$ denotes the vector of covariates measured from the $i$th individual at time $t$ where $d$ is the dimension of the observation, $A_{i,t}$ denotes the treatment assigned to the $i$th individual at time $t$, and $R_{i,t}\in \mathbb{R}$ denotes the $i$th individual's clinical outcome at time $t$. Let $\mathcal{O}$ and $\mathcal{A}$ denote the observation and action space, respectively. We assume $\mathcal{A}$ is a discrete space whereas, $\mathcal{O}$ is a compact subspace of $\mathbb{R}^d$, and the reward is uniformly bounded. The bounded rewards assumption is commonly imposed in the RL literature \citep[see e.g.,][]{fan2020theoretical,li2023q}.

\textbf{\textit{Model}}. We next present the proposed two-way doubly inhomogeneous decision process model. In the RL literature, a common practice is to employ the Markov decision process \citep[MDP,][]{puterman2014markov} to model the data generating process. For simplicity, for now we assume both the observation and reward spaces are discrete, the MDP model essentially requires the reward and future observation to be conditionally independent of the past data history given the current observation-action pair so that the system dynamics are uniquely determined by the following Markov transition function $p$,
\begin{eqnarray}\label{eqn:MDP}
    \prob(O_{i,t+1}=o', R_{i,t}=r|A_{i,t}=a, O_{i,t}=o, \{O_{i,j},A_{i,j},R_{i,j}\}_{1\le j<t})=p(o',r|a,o),
\end{eqnarray}
which is assumed to be doubly homogeneous, i.e., constant over time and population.  

Instead of adopting the MDP model, we propose to use a more general model that relies on two key assumptions. First, we assume the existence of a set of individual- and time-specific latent factors $\{U_i\}_{i=1}^N$ and $\{V_t\}_{t=1}^T$ conditional on which the Markov assumption holds. More specifically, for any $i$ and $t$, we assume
\begin{eqnarray}\label{eqn:assump}
    &&\prob(O_{i,t+1}=o', R_{i,t}=r|U_i=u_i, V_t=v_t, A_{i,t}=a, O_{i,t}=o, \{O_{i,j},A_{i,j},R_{i,j},V_j\}_{1\le j<t})\nonumber\\
    &=&p(o',r|u_i,v_t,a,o).
\end{eqnarray}

\begin{remark}
    Unlike \eqref{eqn:MDP}, the transition function in \eqref{eqn:assump} is both individual- and time-dependent due to the inclusion of $U_i$ and $V_t$. The individual-specific factors can be viewed as certain individual baseline information (e.g., educational background) that does not vary over time whereas the time-specific factors correspond to certain external factors (e.g., holidays) that have common effects on all individuals. 
\end{remark}

\begin{remark}\label{remark:POMDP}
Both $\{U_i\}_{i=1}^N$ and $\{V_t\}_{t=1}^T$ are unobserved in practice, leading to the violation of the Markov assumption. Indeed, the proposed data generating process can be viewed as a special class of partially observable MDPs \citep[POMDPs,][]{sutton2018} where the unobserved factors either do not evolve over time (e.g., $\{U_i\}_{i=1}^N$) or do not vary across individuals (e.g., $\{V_t\}_{t=1}^T$). More generally, one may allow the latent factors to evolve over both time and population. However, this makes the subsequent policy evaluation extremely challenging. In contrast, our proposal decomposes these factors into individual-only and time-only effects, which can be consistently estimated when both $N$ and $T$ diverge to infinity. %We also remark that 
Such latent factor models are widely used in finance \citep{ross1976arbitrage}, economics \citep{bai2002determining} and psychology \citep{bollen2002latent}. %and machine learning \citep{nickel2011three}.
\end{remark}
\begin{remark}
The assumption of discrete rewards and observations is utilized merely to simplify the presentation. Our proposed methodology can be equally applied to scenarios with continuous observation/reward spaces as well. In these cases, we use $p(\bullet,\bullet|U_i,V_t,A_{i,t},O_{i,t})$ to represent the conditional density function of $(O_{i,t+1},R_{i,t})$ given $U_i$, $V_t$, $A_{i,t}$, and $O_{i,t}$.
\end{remark}

Second, we impose an additivity assumption, which requires the transition function $p$ to be additive in $u$, $v$ and $(a,o)$, i.e.,
\begin{eqnarray} \label{eqn:model}
 p(o',r|u_i, v_t, a, o)= \pi_{u} p_{u_i}(o',r|u_i)+\pi_{v} p_{v_t}(o',r|v_t)+\pi_0 p_0(o',r|a,o)
\end{eqnarray}for some non-negative constants $\pi_u$, $\pi_v$, and $\pi_0$ that satisfy $\pi_u+\pi_v+\pi_0=1$ and some unknown conditional probability density (mass) functions $p_u$, $p_v$ and $p_0$. 

The additivity assumption in \eqref{eqn:model} essentially assumes that the transition function corresponds to a mixture of $p_u$, $p_v$ and $p_0$, with the mixing weights given by $\pi_u$, $\pi_v$ and $\pi_0$, respectively. Under the additivity assumption, $p_u$ and $p_v$ correspond to the individual- and time-specific effects, respectively, and are independent of the current observation-action pair. The function $p_0$ corresponds to the main effect shared over time and subjects. Meanwhile, such an additivity assumption can be further relaxed; see Section \ref{sec:extension} for details. 
%. In Section \ref{sec:action-dependant} later, we discuss the setting in which time and subject latent factors are allowed to depend on the action; while in Section \ref{sec:interactive}, we further consider the transition and the reward has the \textbf{\textit{interactive}} effects (factor structure) between time and subject. }

Multiplying both sides of \eqref{eqn:model} by $r$ and integrating with respect to $r$ and $o'$, we obtain
\begin{eqnarray}\label{eqn:reward}
    R_{i,t}=\theta_i + \lambda_t + r_1(a, o)+\varepsilon_{i,t},
\end{eqnarray}
where $\theta_i=\pi_u \int rp_{u_i}(o', r|u_i)drdo'$, $\lambda_t=\pi_v \int rp_{v_t}(o',r|v_t)drdo'$, $r_1(a,o)=\pi_0 \int rp_0(o',r|a,o)drdo'$, and $\varepsilon_{i,t}=R_{i,t}-\Mean (R_{i,t}|A_{i,t}=a,O_{i,t}=o)$ {has conditional mean zero}. Models of this type are referred to as the two-way fixed-effects (2FE) model in the panel data literature \citep[see e.g.,][]{imai2021use}. Nonetheless, our model allows the current treatment to not only affect the immediate outcome, but also impact the future outcomes through its effect on the future observations via the transition function in \eqref{eqn:model}. 

\begin{remark}\label{remark:2FEmodel}
Our additivity assumption \eqref{eqn:model} is motivated by the increased popularity of the fixed-effect models in the panel data literature, due to its interpretability and the ability to account for unobserved variables. As commented by \citet{green2001dirty}, fixed effects regression can scarcely be faulted for being the bearer of bad tidings. Such a model has emerged as a crucial tool assisting researchers in various fields such as medical and political science, facilitating the derivation of scientific conclusions \citep{hotz2011impact,bachhuber2014medical,dwivedi2022counterfactual}. 
%For example, \citet{bachhuber2014medical} studied the association between legal medical marijuana and opioid overdose mortality. They employed 2FE regression and found a statistically significant negative impact on the state-level opioid overdose mortality rates along with the legalization of the medical usage of cannabis. In another influential paper, \citet{hotz2011impact} investigated the influence of state child care regulations on the demand of care in child care markets in USA. Using 2FE regression, they concluded that the implementation of regulations leads to a decrease of the number of center-based child care facilities. This demonstrates the potential of imposing additivity assumption and using 2FE models for decision process modeling.
%modeling the decision process in our setting.  
%On the other hand, when $\pi_0=1$, the proposed model reduces to the standard MDP studied in doubly homogeneous environments. 
\end{remark}

To summarize the data generating process, the latent factors $\{U_i\}_i$ and $\{V_t\}_t$ are sampled prior to all interactions with the environment. For a specific trajectory $i$, at each time point $t$, 
we observe $O_{i,t}$ according to the transition model \eqref{eqn:model}. 
Next, the agent takes an action $A_{i,t}$ 
according to the observed data history and receives an immediate reward $R_{i,t}$ according to \eqref{eqn:model}. 
Finally, the environment transits into the next state, yielding $O_{i,t+1}$. A causal diagram illustrating the data generating procedure can be found in Figure 1 of the supplementary article. 
%Importantly, we focus on the RL settings where the actions are unconfounded in this article. Extensions to settings with unmeasured confounders are left for future research. 
In what follows, we assume the latent factors $\{U_i\}_i$ and $\{V_t\}_t$ are fixed and use $\{u_i\}_i$ and $\{v_t\}_t$ to denote their realizations. Other random variables in the environment will not alter their values.  In the sequel, all the expectations mentioned are implicitly conditional on $\{U_i\}_i$ and $\{V_t\}_t$. %Notably, the policy value of interest is defined as the conditional expectation of the reward given $\{U_i\}_i$ and $\{V_t\}_t$.

\textbf{\textit{Estimands}}. Finally, we define our target estimand of interest. A policy prescribes how an agent acts and makes decisions. Mathematically, it maps the space of observed data history to a probability mass function on the action space, representing the probability that a given individual receives a given treatment at each time point. Throughout this paper, we focus on evaluating \textit{stationary} policies where the action selection probability depends on history only through the current observation and this dependence is stationary over time. More specifically, following a given stationary policy $\pi$, the $i$th individual will receive treatment $a$ with probability $\pi(a|O_{i,t})$. Meanwhile, the proposed method can be extended to evaluate possibly history-dependent policies; see Section A.2 of the supplementary article.

For a target policy $\pi$, we define the following four estimands of interest: (i) the average effect  $\eta^{\pi}:= (NT)^{-1}\sum_{i=1}^N\sum_{t=1}^T \Mean^{\pi} (R_{i,t})$; (ii) the individual-specific effect given the observed initial observation $\eta^{\pi}_i:= T^{-1}\sum_{t=1}^T \Mean^{\pi} (R_{i,t}|O_{i,1})$; %(which is a scalar instead of a function of the initial observation);
(iii) the time-specific effect $\eta^{\pi}_t:= N^{-1}\sum_{i=1}^N\Mean^{\pi} (R_{i,t})$ and (iv) the individual- and time-specific effect $\eta^{\pi}_{i,t}:= \Mean^{\pi} (R_{i,t}|O_{i,1})$. Here, $\E^{\pi}$ means that the expectation is taken by assuming the system dynamics follow the target policy $\pi$. In defining $\eta_{i}^{\pi}$ and $\eta_{i,t}^{\pi}$, we include $O_{i,1}$ in the conditioning set to eliminate their variability resulting from marginalizing over the initial observation distribution. This is reasonable, as the initial observation distribution may no longer be identical across different subjects due to individual heterogeneity, making it impossible to infer consistently from the data. We focus on estimating (iv) $\eta^{\pi}_{i,t}$ in the next two sections, based on which estimators for (i)--(iii) can be easily derived by taking the average over time and/or population.

% \begin{remark}
% In stationary environments with certain mixing conditions \citep[see e.g.,][]{bradley2005basic}, the system will reach its stationary distribution. In that case, it is immediate to see that $\lim_{T} \eta_i^{\pi}=\lim_{t} \eta_{i,t}^{\pi}$ and $\lim_{T} \eta^{\pi}=\lim_{t} \eta_{t}^{\pi}$. Meanwhile, under individual homogeneity, we obtain $\lim_T \eta^{\pi}=\lim_T \eta_i^{\pi}$. 
% %Under individual homogeneity, it is immediate to see that $\eta^{\pi}=\eta_i^{\pi}$ and $\eta_t^{\pi}=\eta_{i,t}^{\pi}$. In addition, when the system enters the stationary observation, we have $\eta^{\pi}=\eta_t^{\pi}$ and $\eta_i^{\pi}=\eta_{i,t}^{\pi}$. 
% As such, the aforementioned four targets are asymptotically the same as in the classical doubly homogeneous environments. 
% \end{remark}

% \begin{remark}
% As commented earlier, we implicitly conditional on the latent factors when taking the expectation with respect to the reward. As such, these values are defined by assuming that we implement the target policy to the exact same population at the exact same time period \LW{This is a good remark but I am also wondering if it will lead to some tough questions from referees} and the target policy will not change the values of these latent factors. It is also interesting to learn the marginal expectation of the reward aggregated over different realizations of the latent factors. However, this is challenging without imposing distributional assumptions on the latent factors. 
% \end{remark}

\section{Model-free OPE}
\label{sec:mdfree}
We now develop model-free methodologies to learn $\eta^{\pi}_{i,t}$: the $i$th subject's average reward at time $t$ given $O_{i,1}$. Model-free methods construct the estimator without directly learning the transition function. Compared to model-based methods which directly learn the transition function to derive the estimator, they are preferred in settings with a large observation space, or where the transition function is highly complicated and can be misspecified. In RL, both model-free and model-based methods have their own unique strengths, and we discuss this point thoroughly in Section A.3 of the supplementary article.

\textbf{\textit{Challenge}}. Before presenting our proposal, we outline the challenges in consistently estimating the policy value. First, existing model-free methods developed in the RL literature \citep[see e.g.,][]{luckett2020estimating,shi2020statistical} focused on learning the long-term reward in a stationary environment. These methods are not applicable to learn the expected reward at a given time with nonstationary transition functions. Second, in the DTR literature, backward induction or dynamic programming is widely employed to evaluate the value function in the sparse reward setting where the reward is obtained at the last stage and all the immediate rewards equal zero. %These methods are 
It is applicable to evaluate $\Mean^{\pi} (R_{i,t})$ in nonstationary environments. Nonetheless, it requires all individual trajectories to follow the same distribution and is thus inapplicable to our setting.

\textbf{\textit{Q-function}}. Our proposal extends the backward induction to the doubly inhomogeneous environments. We first define the following individual- and time-specific Q-function 
%Toward that end, define the Q-function as 
\begin{eqnarray}\label{eqn:Q}
    Q^{\pi}_{i,t_1,t_2}(o,a)=\Mean^{\pi} (R_{i,t_2}|A_{i,t_1}=a, O_{i,t_1}=o), 
\end{eqnarray}
for any $1\le i\le N$ and $1\le t_1\le t_2\le T$. %Our Q-function differs from those in 
To elaborate on this definition, we consider two particular choices of $t_1$. First, when $t_1=t_2$, \eqref{eqn:Q} reduces to the conditional mean of $R_{i,t_2}$ given ($A_{i,t_2}$, $O_{i,t_2}$), which equals $\theta_i+\lambda_t+r_1(A_{i,t},O_{i,t})$ (see Equation (\ref{eqn:reward})) under additivity. Second, when $t_1=1$, it is immediate to see that 
\begin{eqnarray}\label{eqn:etapiitQ1}
\eta^{\pi}_{i,t_2}=\sum_a Q^{\pi}_{i,1,t_2}(O_{i,1},a)\pi(a|O_{i,1}).
\end{eqnarray}
As such, it suffices to learn $Q^{\pi}_{i,1,t}$ to construct estimators for $\eta_{i,t}^{\pi}$. 

\begin{remark}
    In %both the DTR and 
    the RL literature, the Q-function 
    %DTR literature, the Q-function 
    is typically defined as the cumulative reward starting from a given time $t_1$. Our Q-function in \eqref{eqn:Q} differs in that: (i) it is individual-specific where the subscript $i$ encodes its dependence upon the latent factor $u_i$; (ii) it is the conditional mean of the immediate reward at time $t_2$ only instead of the cumulative reward since our objective here lies in evaluating $\Mean^{\pi} (R_{i,t_2})$. 
\end{remark}

%The subscript $i$ encodes the dependence upon 

%From the definition above, it can be seen that $ Q^{\pi}_{i,1,t}(s,a)$ is the expectation of reward $R_t$ assume that action $a$ is taken in a given observation $s$ and thereafter following target policy $\pi$. Hence, our objective lies in estimating $Q^{\pi}_{i,1,t}$ for any $i$ and $t$, the resulting estimator of $\eta^{\pi}_{i,t}$ then can be obtained using $\widehat \eta^{\pi}_{i,t}=\sum_a \widehat Q^{\pi}_{i,1,t}(o_{i,1},a)\pi(a|o_{i,1})$. Next we employ a backward induction procedure to compute the estimator of the defined Q-function. 

\textbf{\textit{Backward induction}}. We propose to use backward induction to compute an estimated Q-function $\widehat{Q}_{i,1,t}^{\pi}$ for $Q_{i,1,t}^{\pi}$ and then plug this estimator into \eqref{eqn:etapiitQ1} to construct the policy value estimator. To begin with, consider the reward function $\{Q^{\pi}_{i,t,t}\}_{i,t}$. 
As shown in \eqref{eqn:reward},
under the two-way fixed-effect model, we have $Q^{\pi}_{i,t,t}(o,a)=r_1(o,a)+\theta_i+\lambda_t$ for any $i$ and $t$. This motivates us to consider the following optimization problem: \begin{eqnarray}\label{eqn:rewardleastsquare}
(\widehat{\btheta},\widehat{\blambda},\widehat{r}_1)=\argmin_{\btheta, \blambda,\, r_1}\sum_{i,t} [R_{i,t}-\theta_i -\lambda_t -r_1(O_{i,t},A_{i,t})]^2, \label{eq:tw}
\end{eqnarray} 
where $\btheta=(\theta_1,\dots,\theta_N)^\top\in \mathbb{R}^{N}$, and $\blambda=(\lambda_1,\dots,\lambda_T)^\top\in \mathbb{R}^{T}$. To guarantee the uniqueness of the solution to \eqref{eqn:rewardleastsquare}, we impose the identifiability constraints $\sum_i \theta_i=\sum_t \lambda_t=0$. There are other constraints one could consider, but they all lead to the same final estimators. %{\color{red}The proposed backward induction algorithm is designed to accommodate both linear and nonlinear function approximation.It can utilize many existing supervised learning algorithms to solve \eqref{eqn:rewardleastsquare}. We provide a concrete proposal based on the method of sieves \citep{grenander1981abstract} in Section \ref{sec:sieve}. Alternatively, one can model $r_1$ via deep neural networks (DNN) and estimate the parameters involved in the DNN as well as $\{\theta_i\}_i$, $\{\lambda_t\}_t$ via e.g., the Adam algorithm \citep{kingma2015adam}.}

We next estimate $\{Q_{i,t-1,t}\}_{i,t}$. According to the Bellman equation, we obtain 
\begin{eqnarray*}
    Q_{i,t-1,t}(O_{i,t-1},A_{i,t-1})=\Mean \Big[\sum_a \pi(a|O_{i,t})Q_{i,t,t}(O_{i,t},a)\big| A_{i,t-1}, O_{i,t-1}\Big].
\end{eqnarray*}Under the additivity assumption, we can similarly obtain a two-way decomposition for $Q_{i,t-1,t}$; see Proposition \ref{prop1} below for a formal statement. This allows us to solve a constrained optimization problem similar  to \eqref{eqn:rewardleastsquare} to estimate $Q_{i,t-1,t}$. We next repeat this procedure to recursively estimate $\{Q_{i,t-2,t}\}_{i,t}$, $\{Q_{i,t-3,t}\}_{i,t}$, $\cdots$,  $\{Q_{i,1,t}\}_{i,t}$ based on the Bellman equation and finally construct the policy value estimator via \eqref{eqn:etapiitQ1}. We summarize our estimating procedure in Algorithm \ref{alg:alg1} below. %and the two-way ANOVA model structure. 
%$Q_{i,t-1,t}(A_{i,t-1},O_{i,t-1})=\Mean [\sum_a \pi(a|O_{i,t})Q_{i,t,t}(a,i,t)]$
%Next, under the given model assumption, according to the Bellman equation, we can similarly show that $Q^{\pi}_{i,t-1,t}$ has a two-way ANOVA model structure. More general, it can be shown that for any $t^*\leq t$, the function $Q^{\pi}_{i,t-t^*,t}(s,a)$ also has a two-way ANOVA model structure (see Proposition \ref{prop1} below). 
% Following \citet{su2006profile} and \citet{lin2014consistent}, we use a  version of least squares dummy variable method to estimate the Q-function. To do so, we introduce dummy variables $\bh_i \in \mathbb{R}^{N}$ for $1 \leq i \leq N$ and $\bg_t \in \mathbb{R}^{T}$ for $1\leq t \leq T$, such that for subject $i$ at time $t$, the $i$th component in $\bh_i$ and the $t$th component in $\bg_t$ is $1$, while all other components in the two vectors are zero. 
The following proposition formally states the two-way structure of these Q-functions. 
\begin{prop}\label{prop1}{For any integer $k<t$, the Q-function
 $Q^{\pi}_{i,t-k+1,t}(o,a)$ satisfies $$Q^{\pi}_{i,t-k+1,t}(o,a)=r^{\pi}_{k}(o,a)+\theta^{\pi}_{k,i}+\lambda^{\pi}_{k,t},%+\varepsilon^{\pi}_{k,i,t},
 $$  where %$r^{\pi}_{k}(o,a)\equiv \E^{\pi}[(r(o_t,a_t)|o_{t-k}=o, a_{t-k}=a]$, 
 $\theta^{\pi}_{k,i}$ and $\lambda^{\pi}_{k,t}$ depend only on $i, k, \pi$ and $t,k, \pi$ respectively. %and $\varepsilon^{\pi}_{t^*,i,t}$ is a mean-zero residual satisfies the Markov assumption. 
}
\end{prop} In what follows, we omit $\pi$ in $r^{\pi}_{k}(o,a), \mbox{ } \theta^{\pi}_{k,i}$, and $\lambda^{\pi}_{k,t}$ when there is no confusion.
%Proposition \ref{prop1} allows us to similarly apply the same strategy to estimate all the Q-functions. This procedure can be sequentially implemented, a pseudocode summarizing our proposed approach to estimate $\eta^{\pi}_{i,t^*}$ for a given $t^*$ can be found in Algorithm \ref{alg:alg1}. We remark that all the values of $\widehat \eta^{\pi}_{i,t}$ such that $t\leq t^*$, can also be obtained in the process of estimating $\eta^{\pi}_{i,t^*}$ (see Algorithm \ref{alg:alg1}). For brevity, in what follows, we will omit subscript $\pi$ for $r^{\pi}_{t^*}(o,a), \mbox{ } \theta^{\pi}_{t^*,i}$, and $\lambda^{\pi}_{t^*,t}$ when there is no confusion.

\spacingset{1.0}
\begin{algorithm}[t] 
	\caption{Pseudocode for Estimating 
  $\eta_{i,t^*}^{\pi}$.}
	\begin{algorithmic}[1]
		\small
		\Function{}{$ R_{i,t},A_{i,t},O_{i,t} $ for $1\le i\le N, 1\le t\le T$}
  \State Set iteration counter $k \gets 1$.
        \State Solve $ (\widehat\theta_i, \widehat\lambda_t, \widehat r_1)=\argmin_{\theta_i, \lambda_t, r_1}\sum_{i,t} [R_{i,t}-\theta_i-\lambda_t-r_1(O_{i,t}, A_{i,t})]^2.$
        \State Compute $\widehat Q^{\pi}_{i,t,t}(O_{i,t},a)=\widehat r_1(O_{i,t}, a)+\widehat\theta_i+\widehat\lambda_t$, for all $i$, $t$, and $a\in \mathcal{A}$.
        \Repeat
        \State $k \gets k + 1$
        \State For all $i$ and $t\geq k$, solve \begin{gather*}
            \min_{\theta_{k,i}, \lambda_{k,t}, r_k}\sum_{i,t} \;\left[\sum_{a\in \mathcal{A}}\pi(a|O_{i,t-k+2}) \widehat Q^{\pi}_{i,t-k+2,t}(O_{i,t-k+2},a)-\theta_{k,i}-\lambda_{k,t}-r_k(O_{i,t-k+1},A_{i,t-k+1})\right]^2.
        \end{gather*} 
        \State Compute $\widehat Q^{\pi}_{i,t-k+1,t}(O_{i,t-k+1},a)=\widehat r_k(O_{i,t-k+1}, a)+\widehat\theta_{k,i}+\widehat\lambda_{k,t}$, for $a\in \mathcal{A}$.
        \Until{$k=t^*$}
       \State $\widehat \eta^{\pi}_{i,t^*}=\sum_a \widehat Q^{\pi}_{i,1,t^*}(O_{i,1},a)\pi(a|O_{i,1})$
		\EndFunction
\end{algorithmic}\label{alg:alg1}
\end{algorithm}

\spacingset{1.7}

To conclude this section, we draw a comparison with the classical backward induction in the DTR literature \citep{Murphy,gest}. First, the classical backward induction algorithm aims to learn the Q-function under an optimal policy and derive the optimal policy as the greedy policy with respect to the estimated Q-function. To the contrary, the proposed algorithm learns the Q-function under a fixed target policy for the purpose of policy evaluation. Second, classical backward induction requires the computation of the Q-function recursively from time $t$ till the beginning. However, it is worth mentioning that %under certain mixing conditions, %for our proposed methods, 
%the proposed main effect $r_{t^*}$  %, individual-specific effects $\theta_{t^*,i}$ and time-specific effects  
%and the two-way fixed effects $\theta_{t^*,i}$, $\lambda_{t^*,t}$ will converge
%exponentially fast with respect to the lag $t^*$ (see the proof of Proposition \ref{prop1} in the Appendix for details). 
our estimated Q-function converges exponentially fast to a constant function with respect to the lag $k$
(see Section B.2.2 of the supplementary material). We refer to this phenomenon as Q-function degeneracy. As such, early stopping can be potentially employed in Algorithm \ref{alg:alg1} to speed up the computation.  %there is no need to iterate $t^*$ many steps to obtain the estimates of $\eta_{i,t^*}^{\pi}$. This is one of the strengths of our proposed method.

%In addition, our 
Finally, the proposed backward induction allows us to to efficiently borrow information under the additivity assumption. Specifically, during each iteration, we pull all the relevant data together to estimate the Q-function. This allows us to consistently estimate the main effect (shared by all observations) at a rate of $(NT)^{-\alpha}$ which depends on both $N$ and $T$, and the exponent $0<\alpha\le 1/2$ depends on the nonparametric methods being used to solve the optimization problem. Meanwhile, the %individual- and time-specific 
two-way fixed effects $\theta$s and $\lambda$s converge at $T^{-1/2}$ and $N^{-1/2}$ respectively, up to some logarithmic factors. To the contrary, the estimator obtained via the classical backward induction typically converges at a rate of $N^{-\alpha'}$ for some $0<\alpha'\le 1/2$ in individual-homogeneous and history-dependent\footnote{The transition function depends on the entire history instead of the current observation-action pair.} environments. 
%where the exponent $\alpha$ depends on the  
%the main effect is estimated 
%use pre-collected data, compared to the standard backward induction. To see this, noted that in the process of estimating the Q-function $Q^{\pi}_{i,t,t}$, our proposed approach permits us to utilize data from all trajectories across all time periods to estimate the reward function. In contrast, the standard backward induction only takes advantage of the data at the specific time point. Moreover, the standard backward induction does not take the subject heterogeneity into account.

\subsection{A Linear Sieve Estimator for Two-way Fixed Effects Model}\label{sec:sieve}

%\bian{
\textbf{\textit{Notation}}. 
Given arbitrary $\{x_{i,t}\}_{1\leq i\leq N, 1\leq t\leq T}$,
%we construct the vector 
let $\boldsymbol{x}\in \mathbb{R}^{NT}$ %such that its 
denote the vector whose $((t-1)N+i)$-th element equals $x_{i,t}$. That is, $\boldsymbol x$ is constructed by stacking the $N$ elements at the first time point, followed by the $N$ elements at the second time point, and continuing in this manner until the $N$ elements from the final time point are included, i.e.,
$$\boldsymbol x=(x_{1,1},x_{2,1},\dots,x_{N,1},x_{1,2},\dots,x_{N-1,T},x_{N,T})^\top.$$ Similarly, given a set of vectors $\{\boldsymbol{x}_{i,t}\}_{i,t}$, let $\boldsymbol X$ denote the matrix whose $((t-1)N+i)$th row equals $\boldsymbol{x}_{i,t}$. To implement Algorithm \ref{alg:alg1}, we need to solve two-way fixed effects models repeatedly for value function estimation. 
To simplify the presentation, we focus on %the one time interval two-way fixed effects model , i.e., 
the estimation of $Q^{\pi}_{i,t,t}(O_{i,t},a_{i,t})$ (see Equation \eqref{eq:tw}). We propose to approximate the %non-parametric part 
main effect function $r_1(o,a)$ %in the Q-function 
using linear sieves \citep{huang1998projection,chen2015optimal}. Under mild conditions, there exists a set of vectors $\{\bbeta^*_a\}$ such that the approximation error is negligible, i.e., $\sup_{o, a} |r_1(o,a)-\bPhi_L(o)^\top\bbeta^*_a|=O(L^{-p/d})$, where $\bPhi_L(o)$ is a vector consisting of $L$ sieve basis functions, e.g., splines or wavelet bases, and $p>0$ measures the smoothness of the system dynamics; see Section B.2.1 for more details. 
For simplicity, we now focus on the binary action space setting, in which $\mathcal{A}=\{0,1\}$.

The two-way fixed effects model in \eqref{eqn:model} can be represented in the following matrix form: 
$\bR=\bB\balpha+\bM+\boldsymbol\varepsilon,$ where $\bR=(R_{1,1},R_{2,1},\dots,R_{N,1},R_{1,2},\dots,R_{N-1,T},R_{N,T})^\top \in \mathbb R^{NT} $, 
$\balpha=(\btheta^\top,\blambda^\top)^\top$,
$\bB=(\1_T \otimes \bI_N, \bI_T \otimes \1_N)\in \mathbb R^{NT\times (N+T)} $ is the design matrix, $\bI_N$ is a $N \times N$ identity matrix, $\1_T$ is a vector of length $T$ with all elements one, $\bM=(r_1(O_{1,1},A_{1,1}),r_1(O_{2,1},A_{2,1}),\dots,r_1(O_{N,1},A_{N,1}),r_1(O_{1,2},A_{1,2}),\cdots,r_1(O_{N,T},A_{N,T}))^\top\in \mathbb R^{NT} $, and $\otimes$ is the Kronecker product. In what follows, we will omit the indices of these matrices and vectors
%index of identity matrix and the vector of all ones 
when there is no confusion. Let $\bPhi_{i,t}=((1-A_{i,t})\bPhi_L^\top(O_{i,t}),A_{i,t}\bPhi_L^\top(O_{i,t}) )^\top$, and let $\bPhi$ be the $\mathbb R^{NT \times 2L}$ matrix %such that 
$$(\bPhi_{1,1}^\top,\bPhi_{2,1}^\top,\dots,\bPhi_{N,1}^\top,\bPhi_{1,2}^\top,\dots,\bPhi_{N-1,T}^\top,\bPhi_{N,T}^\top)^\top.$$ By the Frisch–Waugh–Lovell theorem \citep{frisch1933partial,lovell1963seasonal}, a closed-form estimator of $\bbeta=(\bbeta_0^\top,\bbeta_1^\top)^\top$ can be obtained accordingly: \begin{gather} \label{eq3}
    \widehat\bbeta=(\bPhi^\top(\bI-\bP)\bPhi)^{-1}\bPhi^\top(\bI-\bP) \bR,
\end{gather} where $\bP$ is the projection matrix: $\bP=\bB(\bB^\top \bB)^+\bB^\top$, and $(\bB^\top \bB)^+$ is the Moore–Penrose inverse of the matrix $\bB^\top \bB$. Given $\widehat{\bbeta}$, the estimator for the main effect function $r_1(O_{i,t},A_{i,t})$ (denoted by $\widehat{r}_1$) can then be obtained, %then we can accordingly calculate 
based on which the fixed effects can be estimated. Specifically, under the constraints that $\sum_{i=1}^N \theta_i=\sum_{t=1}^T \lambda_t=0$, we have $$\widehat \theta_i=T^{-1} \sum_{t=1}^T (R_{i,t}-\widehat r_1(O_{i,t},A_{i,t})),\mbox{ and }\widehat \lambda_t=N^{-1} \sum_{i=1}^N (R_{i,t}-\widehat r_1(O_{i,t},A_{i,t})).$$ The resulting estimated Q-function %at the last time interval 
is given by $\widehat Q^{\pi}_{i,t,t}(o,a)=\widehat\theta_i+\widehat\lambda_t+ \bPhi_L(o)^\top \widehat\bbeta_a$. 

% \begin{remark}
%     When $N$ or $T$ is large, solving the two-way fixed effects models directly can be computationally burdensome. An alternative approach is to separate the estimation procedure into two tasks so that the main effect can be computed first, followed by the estimation of time and individual-specific effects. %As such, it 
%     This avoids inverting a large $(N+T+2L)\times (N+T+2L)$ square matrix, making it computationally efficient (see Algorithm 2 in Section \ref{alg:alg2} of the supplementary article).
% \end{remark}

\subsection{Beyond the additivity assumption} \label{sec:extension} 

In this section, we discuss two extensions of the additivity assumption \eqref{eqn:model}. %In Section \ref{sec:action-dependant} below, we 
This first extension allows the time- and individual-specific latent factors to additionally depend on the action, whereas the second extension permits the %observation transition and the reward 
system dynamics to incorporate interactive effects between time and individual. %. In Section \ref{sec:interactive}, we %further consider the setting in which 
%allow the transition and reward functions to have the \textbf{\textit{interactive}} effects %(factor structure)  between time and subject.

\textbf{\textit{Action-dependant time- and individual-specific effects}}. 
We first consider the following relaxation of the additivity assumption:
\begin{eqnarray*} 
 p(o',r|u_i, v_t, a, o)= \pi_{u} p_{u_i,a}(o',r|u_i)+\pi_{v} p_{v_t,a}(o',r|v_t)+\pi_0 p_0(o',r|a,o),
\end{eqnarray*} 
such that $\pi_u+\pi_v+\pi_0=1$. That is, we now allow both  the $p_{u_i}(o',r|u_i)$ and $p_{v_t}(o',r|v_t)$ in \eqref{eqn:assump} to depend on the action $a$, and now the Q-function $Q^{\pi}_{i,t-k+1,t}(o,a)$ satisfies $$Q^{\pi}_{i,t-k+1,t}(o,a)=r^{\pi}_{k}(o,a)+\theta^{\pi}_{k,i}(a)+\lambda^{\pi}_{k,t}(a), $$  where $\theta^{\pi}_{k,i}(a)$ and $\lambda^{\pi}_{k,t}(a)$ %are two constants depend only on $i, k, \pi,a$ and $t,k, \pi,a$ respectively. 
are action-dependent. Since $a$ is binary, each iteration requires to estimate $2N+2T_k$ fixed effects. The proposed approach can be easily extended to solve this new problem without extra complications. We omit the details to save space. 

\textit{\textbf{Interactive time- and individual-specific effects}}. The second extension is motivated by the factor model, which is extensively used in the panel data literature to relax the additivity assumption \citep{bai2002determining}. In our setup, consider the reward regression model in \eqref{eqn:reward}. The factor model replaces the additive terms $\theta_i + \lambda_t$ in \eqref{eqn:reward} with an interaction term $\bgamma_i^\top \balpha_t$, resulting in 
\begin{align}\label{eqn:rewardfactor}
    R_{i,t}=\bgamma_i^\top \balpha_t + r_1(A_{i,t}, O_{i,t})+\varepsilon_{i,t},
\end{align}
where $\bgamma_i \in \mathbb R^{h}$ and $\balpha_t \in \mathbb R^{h}$
denote the vectors of unobserved common time- and individual-specific factors, respectively. By definition, it covers the additive model as a special case by setting $h=2$, $\bgamma_i=(1, \theta_i)^\top$ and $\balpha_t=(\lambda_t,1)^\top$. 

Combining \eqref{eqn:rewardfactor} together with a completeness assumption, which requires functions in the form of the right-hand-side of \eqref{eqn:rewardfactor} to be closed under the Bellman operator (see Assumption \ref{asmp: completeness} in Section \ref{sec:theory}), we can show that $Q_{i,t-k+1,t}$ maintains a factor structure for any $k < t$. Similar to Algorithm \ref{alg:alg1}, backward induction remains applicable for estimating $\eta_{i,t}^\pi$.
%Moreover, if we further assume Bellman completeness assumption in the sense that 
% \begin{gather*} 
% p_{i,t+1}(O_{i,t+1}|o,a)=\left[1-\widetilde\bgamma_{i}^\top \widetilde\balpha_{t}\right]p_0(O_{i,t+1}|o,a)+\widetilde\bgamma_{i}^\top \widetilde\balpha_{t},
% \end{gather*} for $0<\widetilde\bgamma_{i}^\top \widetilde\balpha_{t}<1$ so that $p_{i,t+1}(O_{i,t+1}|o,a)$ is a valid transition, and $\widetilde\bgamma_{i,j}\in \mathbb R^{h_j}$ and $\widetilde\balpha_{t,j} \in \mathbb R^{h_j}$ denotes the unknown loadings and unobserved common factors for $O_{i,t+1,j}$, respectively. 
%using previously established framework along with factor analysis techniques \citep{bai2002determining}, one can still estimate $\eta_{i,t}^\pi$ through dynamic programming. 

Finally, we provide a model diagnostic procedure in Section A.1 in the supplementary article to assess the additivity assumption. Specifically, we tackle the model selection problem of determining whether the additive or interactive model better fits the data. Therein, we apply this procedure across various of synthetic 
environments to demonstrate its effectiveness.
 \vspace{-2.5em}

\section{Model-based OPE}\label{sec:mdbased} 

In this section, we develop model-based methods that derive the off-policy value estimator by learning the system dynamics. Recall that under the additivity assumption, 
\begin{eqnarray*}
	R_{i,t}=\theta_i+\lambda_t+r_1(O_{i,t}, A_{i,t})+\varepsilon_{i,t}.
\end{eqnarray*}
As we discussed in Section \ref{sec:mdfree}, the main effect %reward function 
$r_1$, as well as the individual- and time-specific effects can be estimated by solving the following optimization problem, \begin{eqnarray*}
	\argmin_{\{\theta_i\}_i, \{\lambda_t\}_t, r_1}\sum_{i,t} [R_{i,t}-\theta_i-\lambda_t-r_1(O_{i,t}, A_{i,t})]^2.
\end{eqnarray*} 
In addition, we need to estimate the mixing probabilities $\pi_u$, $\pi_v$, $\pi_0$ as well as the distribution functions $p_{u_i}(o'|u_i)$, $p_{v_t}(o'|v_t)$, $p_0(o'|a,o)$, obtained by marginalizing over $p_{u_i}(o',r|u_i)$, $p_{v_t}(o',r|v_t)$, $p_0(o',r|a,o)$ in Equation \eqref{eqn:model}. 

Given these estimators, we employ a simulation-based method to construct the policy value. To be more specific, based on the estimated transition function, we simulate an observation $O_{i,2}^*$ based on the observed $O_{i,1}$ under the target policy $\pi$. %and the estimated transition function. 
%Then given any 
We next sequentially simulate a sequence of observations $\{O_{i,t}^*\}_{t}$ under $\pi$ %, we 
and compute the estimated reward $\pi(a|O_{i,t}^*)(\widehat{r}_1(O_{i,t}^*, a)+\widehat\theta_i+\widehat\lambda_t)$. %then generate the simulated observations at the next time point using the mixture model; and 
Finally, we repeat this procedure sufficiently many times and average all the estimated rewards across different simulations. 
%until the end. 

\textbf{\textit{Likelihood}}. It remains to estimate $p_u$, $p_v$, $p_0$ and $\pi_u$, $\pi_v$, $\pi_0$. Given the latent factors, the likelihood function is proportional to the following, 
\begin{align}
    \notag &\prod_{i=1}^N\prod_{t=2}^T p(O_{i,t}|u_i, v_{t-1}, A_{i,t-1}, O_{i,t-1};\bTheta)\\
    =&\prod_{i=1}^N\prod_{t=2}^T [\pi_u p_{u_i}(O_{i,t}|u_i;\bTheta_{u})+ \pi_v p_{v_t}(O_{i,t}|v_{t-1},\bTheta_{v})+\pi_0 p_0(O_{i,t}|A_{i,t-1}, O_{i,t-1};\bTheta_{0})], \label{eq:likelihood}
\end{align} where we parameterize the transition model
 by $\bTheta=\{\pi_0,\pi_u,\pi_v,\bTheta_{0},\bTheta_{v},\bTheta_{u}\}$, and $\bTheta_{0},\bTheta_{v},\bTheta_{u}$ are the parameters corresponding to models $p_0,p_v$ and $p_u$ respectively.
 
We introduce a latent variable $Z_{i,t}\in \{0,1,2\}$ such that $Z_{i,t}=0$, if $O_{i,t}$ is generated by $p_0$, $Z_{i,t}=1$, if $O_{i,t}$ is generated by $p_{v}$, and $Z_{i,t}=2$ otherwise. Directly maximizing the likelihood in %Expression 
 \eqref{eq:likelihood} %can be 
 is challenging, since  it requires to marginalize over $Z_{i,t}$. %To address issues related to latent variables, variational auto-encoder \citep{kingma2014auto} and expectation–maximization (EM) algorithm \citep{dempster1977maximum,wu1983convergence,meng1997algorithm} can be used. Below we 
 We propose to use the EM \citep{dempster1977maximum} algorithm for parameter estimation. The EM algorithm recursively alternates between an E-step for computing conditional expectation and an M-step for maximizing the likelihood. We detail the two steps below. 

%By Expression 
\textbf{\textit{E-step}}. Similar to \eqref{eq:likelihood}, the complete log-likelihood involving $\{O_{i,t}\}_{i,t}$ and $\{Z_{i,t}\}_{i,t}$ is 
\begin{align}\label{eqn:completelikelihood}
    \notag & l(\boldsymbol O,\boldsymbol Z|\boldsymbol U, \boldsymbol V,\boldsymbol A;\bTheta)\propto 
     \sum_{i=1}^N\sum_{t=2}^T \log p(O_{i,t}|Z_{i,t},u_i, v_{t-1}, A_{i,t-1}, O_{i,t-1};\bTheta) p(Z_{i,t};\bTheta)\\
     \notag =&\sum_{i=1}^N\sum_{t=2}^T [\mathbb{I}(Z_{i,t}=0)\log (\pi_0 p_0(O_{i,t}|A_{i,t-1},O_{i,t-1};\bTheta_{0}))\\+&\mathbb{I}(Z_{i,t}=1)\log (\pi_v p_{v_t}(O_{i,t}|v_t;\bTheta_{v})) +\mathbb{I}(Z_{i,t}=2)\log (\pi_u p_{u_i}(O_{i,t}|u_i;\bTheta_{u}))].
\end{align} Given a current estimate of $\bTheta$, say $\widetilde\bTheta$, define $\Gamma(\bTheta|\widetilde\bTheta)$ as the expected value of $l(\boldsymbol O,\boldsymbol Z|\boldsymbol U, \boldsymbol V,\boldsymbol A;\bTheta)$ with respect to the currently estimated conditional distribution function $p(\boldsymbol Z|\boldsymbol U,\boldsymbol V,\boldsymbol O,\boldsymbol A;\widetilde\bTheta)$. % and the current guess of the parameter $\widetilde\bTheta$, 
%i.e., $\Gamma(\bTheta|\widetilde\bTheta)=\E_{Z\sim p(\boldsymbol Z|\boldsymbol U,\boldsymbol V,\boldsymbol O,\boldsymbol A;\widetilde\bTheta)}l(\boldsymbol O,\boldsymbol Z|\boldsymbol U,\boldsymbol V,\boldsymbol A;\bTheta)$. 
We aim to calculate $\Gamma$ in this step. It follows from \eqref{eqn:completelikelihood} that %can be shown that 
\begin{align}\label{eqn:Gammatheta}
\begin{split}
    \Gamma(\bTheta|\widetilde\bTheta)=
    &\sum_{i=1}^N\sum_{t=2}^T [p(Z_{i,t}=0|O_{i,t},A_{i,t-1},O_{i,t-1};\widetilde\bTheta_0)\log (\pi_0 p_0(O_{i,t}|A_{i,t-1},O_{i,t-1};\bTheta_{0}))\\
    &+p(Z_{i,t}=1|O_{i,t},v_t;\widetilde\bTheta_{v})\log (\pi_v p_{v_t}(O_{i,t}|v_t;\bTheta_{v}))\\
    &+p(Z_{i,t}=2|O_{i,t},u_i;\widetilde\bTheta_{u})\log (\pi_u p_{u_i}(O_{i,t}|u_i;\bTheta_{u}))].
\end{split}    
\end{align} 
\textbf{\textit{M-step}}. We aim to update the model parameter $\bTheta_{new}$ that maximizes $\Gamma(\bTheta|\widetilde\bTheta)$ with respect to $\bTheta$, i.e., $\bTheta_{new}=\argmax_{\bTheta} \Gamma(\bTheta|\widetilde\bTheta)$. It follows from \eqref{eqn:Gammatheta} that
\begin{eqnarray*}
    \pi_{0,new}&=&\frac{1}{N(T-1)} \sum_{i=1}^N\sum_{t=2}^T p(Z_{i,t}=0|O_{i,t},A_{i,t-1},O_{i,t-1};\widetilde\bTheta_0),\\
    \pi_{v,new}&=&\frac{1}{N(T-1)} \sum_{i=1}^N\sum_{t=2}^T p(Z_{i,t}=1|O_{i,t},v_t;\widetilde\bTheta_{v}),\\
    \pi_{u,new}&=&\frac{1}{N(T-1)} \sum_{i=1}^N\sum_{t=2}^T p(Z_{i,t}=2|O_{i,t},u_i;\widetilde\bTheta_{u}).
\end{eqnarray*}
The rest of the parameters can be updated using any derivative-based (e.g., quasi-Newton) or derivative-free (e.g., Nelder-Mead) algorithm. Our final estimator is obtained by repeating the E-step and the M-step until convergence.  %Iterate the E-step described above and the M-step until convergence, this gives us the EM algorithm. 

\textbf{\textit{Choice of the parametric family}}. In our implementation, when the observation is continuous, we %could 
posit normal distribution functions for %distribution functions 
$p_{u_i}(o'|u_i)$, $p_{v_t}(o'|v_t)$, $p_0(o'|a,o)$, i.e., $p_{u_i}(o'|u_i)=\phi(o';\mu_{u_i},\Sigma_{u_i})$, $p_{u_i}(o'|v_t)=\phi(o';\mu_{v_t},\Sigma_{v_t})$ and $p_0(o'|a,o)=\phi(o';\mu_0(a,o),\Sigma_0(a,o))$ where $\phi(\bullet;\mu,\Sigma)$ denotes a $d$-dimensional multivariate normal density function with mean vector $\mu$ and covariance matrix $\Sigma$. 
We further use a linear model for the mean function $\mu_0$, i.e., $\mu_0(a,o)=\Lambda o +\psi a$ and a constant model for the covariance function, i.e., $\Sigma_0(a,o)=\Sigma_0$ for any $a$ and $o$. 
%and it is sufficient to estimate their expectations $\mu_{u_i}$, $ \mu_{v_t},$ and covariance matrices $\Sigma_{u_i}$ and $\Sigma_{v_t}$. As for $p_0(o'|o,a)$, we could use a linear approximation to estimate its conditional mean, i.e., $\E(O'|A,O)=\Lambda O +\psi A$, where $\Lambda\in \mathbb{R}^{d\times d}$, and $\psi \in \mathbb{R} $. We further assume that $\Cov(O'|A,O)=\Sigma_0 \in \mathbb{R}^{d\times d}$, i.e, the conditional covariance matrix is constant. 
As such, the set of parameters $\bTheta$ can be summarized by $\{\pi_0,\pi_{v},\pi_{u},\{\mu_{u_i}\}_i,\{\Sigma_{u_i}\}_i,\{\mu_{v_t}\}_t,\{\Sigma_{v_t}\}_t,\Lambda, \psi,\Sigma_0\}$.

% In what follows, we describe the estimation procedure for the observation-action transition function based on a posited Gaussian mixture model. 

 \section{Theoretical Results}\label{sec:theory}
 
In this section, we focus on investigating the theoretical properties of our proposed model-free estimators. 
Consistencies and convergence rates of the model-based estimators can be established based on existing analyses of EM algorithms \citep[see e.g.,][]{wu1983convergence,balakrishnan2017statistical} and we omit the details to save space. 

\textbf{\textit{Summary}}. We begin with a summary of our theoretical results. Theorem \ref{thm:rateofconvergence} presents the convergence rates of the proposed value estimators. In particular, for a sufficiently large $L$, we show that the estimated average effect $\widehat{\eta}^{\pi}$, individual-specific effect $\widehat{\eta}_i^{\pi}$, time-specific effect $\widehat{\eta}_t^{\pi}$ and individual- and time-specific $\widehat{\eta}_{i,t}^{\pi}$ converge at a rate of $(NT)^{-1/2}$, $T^{-1/2}$, $N^{-1/2}$ and $\min^{-1/2}(N,T)$, respectively, up to some logarithmic factors. Theorem \ref{thm:asymnormal} establishes the limiting distributions of these estimators. We next impose some technical assumptions. %Recall that $p_0$, $p_u$ and $p_v$ denote the transition function for the main effect and the fixed 
%To estimate the Q-function accurately, we first need certain conditions on the density transition and reward functions. Let $\Sigma(p,C)$ denote the Hölder space of smoothness $p >0$ functions for some constant $C$ (see Appendix for the definition of Hölder class of functions). 

% \begin{asmp}[H{\"o}lder smoothness]\label{asmp:smooth}{
% Assume that there exist some constants $p$ and $C$, such that for any $a\in \mathcal{A}$ and $o'\in \mathcal{O}$, both the function $r_1(\cdot,a)$ and $p_0(o'|\cdot,a)$ belong to the class of $p$-smooth functions $\Lambda(p,C)$; see Appendix \ref{sec:proofsmooth} for the detailed definition.
% }\end{asmp}

\begin{asmp}[Realizability] \label{asmp: realizability}
    Assume that there exist some constants $p$ and $C$, such that for any $a\in \mathcal{A}$, the reward function $r_1(\cdot,a) \in \Lambda(p,C) $, where $\Lambda(p,C)$ is the H{\"o}lder class with the smoothness parameter $p$ (see Section B.2.1 for the definition).
\end{asmp}

\begin{asmp}[Completeness] \label{asmp: completeness}
    For any function $g$ such that $g(\cdot,a) \in \Lambda(p,C)$ for all action $a$, 
    %Assume that there exist some constants $p$ and $C$, such that for any $a\in \mathcal{A}$ and function $g(\cdot,a) \in \Lambda(p,C) $, we have 
    $\mathcal{B}^\pi g(\cdot,a)\in \Lambda(p,C)$ where  $\mathcal{B}^\pi$ denotes the Bellman operator, i.e., $(\mathcal{B}^\pi g)(o,a)=\sum_{a'}\E_{O'\sim p_0(O'|o,a)}[ \pi(a'|O')g(O',a')].$
\end{asmp}

\begin{asmp}[Basis functions]\label{asmp:basis}{
(i) $\sup_{o} \|\bPhi_L(o)\|_2=O(\sqrt{L})$ and $\lambda_{\max}[\int_{o\in \mathcal{O}} \bPhi_L(o) \bPhi^\top_L(o) do]=O(1)$; (ii) For any $C>0$, $\sup_{f\in \Lambda(p, C)} \inf_{\beta\in \mathbb{R}^L} \sup_o |\Phi^\top_L(o) \beta-f(o)|=O(L^{-p/d})$; (iii) $L\ll \min(N,T)/\log(NT)$; (iv) $NT\ll L^{2p/d}$.  
}\end{asmp}

\begin{asmp}[System dynamics]\label{asmp:transition}{
(i) Assume that there exist random errors $\{e_{i,t}\}_{i,t}$ that are i.i.d copies of $E$ such that the future observation $O_{i,t+1}$ can be represented as $\kappa(O_{i,t}, A_{i,t}, u_i, v_t, e_{i,t})$ for some function $\kappa$ that satisfies
\begin{eqnarray*}
    \sup_{a,u,v} \E \norm{\kappa(o,a,u,v,E)-\kappa(o',a,u,v,E)}_2 \leq q \norm{o-o'}_2,\\\sup_{o,a}\norm{\kappa(o,a,u,v,E)-\kappa(o,a,u,v,E')}_2 =O(\norm{E-E'}_2),
\end{eqnarray*}
for some $0\le q<1$; (ii) each element of the error $E$ vector has sub-exponential tail, i.e., $\max_j\Mean \exp(t |E_j|)<\infty$ for some $t>0$, where $E_j$ denotes the $j$th element of $E$; (iii) the reward function $r_1$, the density functions $p_0$, $p_u$,  $p_v$ and $N^{-1} \sum_{i=1}^N p_{O_{i,1}}$ ($p_{O_{i,1}}$ denotes the density function of $O_{i,1}$) are uniformly bounded. 
(iv) there exists some constant $c\geq 1$ such that $\E(\sum_{k=1}^T\varepsilon_{k,i,t}^2|O_{i,t}=o,A_{i,t}=a)>c^{-1}$, for any $i$, $t$, $o \in \mathcal{O}$, $a \in \mathcal{A}$, where $\varepsilon_{k,i,t}$ denotes the Bellman error defined in Section B.3 of the supplementary article. 
}\end{asmp}

\begin{asmp}[Stability]\label{asmp:matrix}{For any backward step $k$ %i.e., 
(the $k$th iteration in Algorithm \ref{alg:alg1}), 
\begin{gather}
  \notag \lambda_{\min}[\E(\bPhi_k^\top \bS_k\bPhi_{k-1}^{new})] \geq (NT)\rho_0 \mbox{ and }
    \norm{[\E(\bPhi_k^\top \bS_k\bPhi_k)]^{-1} \E(\bPhi_k^\top \bS_k\bPhi_{k-1}^{new})}_2\le \rho_1,  \label{eq:irrep}
\end{gather} for some constants $\rho_0>0$ and $0<\rho_1<1$, where $\bPhi_{k}$ is the matrix consisting of the first $N(T-k+1)$ rows of matrix $\bPhi$, $\bS_k$ and $\bB_k$ are the residual maker matrix and the design matrix for the fixed effects at step $k$ respectively (%similarly, $\bS_k$ and $\bB_k$ is the matrix consisting of the first $N(T-k+1)$ rows of matrix $\bS$ and $\bB$ respectively, 
see Section B.2.4.5 of the supplementary article for the detailed formulation), and the matrix $\bPhi_k^{new}$ is a variant of $\bPhi_k$, with its detailed definition provided in Section B.2.4.5 of the supplementary article. %the design matrix such that 
%\begin{eqnarray}
%\bPhi^{new}_{k,i,t}=\frac{1}{T-k+1}\sum_{t=1}^{T-k+1} \bPhi_{k,i,t}+\frac{1}{N}\sum_{i=1}^N \bPhi_{k,i,t}+\sum_{a \in \mathcal{A}}  \pi(a|O_{k,i,t}) \bPhi_{k,i,t}. \label{Phinew}
%\end{eqnarray}
}\end{asmp}

The realizability and the Bellman completeness assumptions are commonly imposed in the RL literature \citep[see e.g.,][]{chen2019information,uehara2022review}. Realizability essentially requires the H{\"o}lder class to be sufficiently rich to contain $r_1$. The Bellman completeness requires the H{\"o}lder class to be ``complete" in the sense that it remains closed under the Bellman operator. It holds automatically when the transition function belong to the H{\"o}lder class as well. The H{\"o}lder smoothness assumption is frequently imposed in the sieve estimation literature \citep{huang1998projection,chen2015optimal}. It has seen increasing adoption in the RL literature as well \citep{fan2020theoretical,chen2022well,shi2020statistical}. %Additionally, recall that $r_1$ and $p_0$ in Assumptions \ref{asmp: realizability} and \ref{asmp: completeness} represent the main effects in the reward and transition functions. Our proposed two-way model simplifies matters in a doubly inhomogeneous environment, allowing these assumptions to be applied solely to the main components. 
Assumption \ref{asmp:basis}(i) and (ii) are automatically satisfied when tensor product B-spline or wavelet bases is used; see Section 6 of \citet{chen2015optimal} for a review of these basis functions. Assumption \ref{asmp:basis}(i) upper bounds the $\ell_2$ norm of the sieve basis vector and the maximum eigenvalue of $\int_{o\in \mathcal{O}} \bPhi_L(o) \bPhi^\top_L(o) do$ whereas Assumption \ref{asmp:basis}(ii) upper bounds the approximation error of the sieve estimator. Assumption \ref{asmp:basis}(iii) upper bounds the number of basis functions and is to guarantee the consistency of the estimator. Assumption \ref{asmp:basis}(iv) lower bounds the number of basis functions,  requiring the bias of our estimator to be much smaller than its standard deviation so as to establish its asymptotic normality. 

Assumption \ref{asmp:transition}(i) requires $\kappa$ to be Lipschitz continuous. Assumption \ref{asmp:transition}(ii) requires the error distribution to possess the sub-exponential tail. These two conditions allow us to establish %the concentration inequality in the presence of a non-stationary transition function, which is required to prove the error bound of the resulting value estimator (see proof of Theorem \ref{thm:rateofconvergence} in the Appendix). %It is also posited in \citep{li2022reinforcement} in order to derive a matrix Bernstein inequality.
concentration inequalities in doubly inhomogeneous environments. Under the additivity assumption  \eqref{eqn:model}, there exist functions $\kappa_0$, $\kappa_u$, $\kappa_v$ and random errors $E_0$, $E_u$ and $E_v$ such that $\kappa(o,a,u,v,E)\stackrel{d}=\mathbb{I}(Z=0)\kappa_0(o,a,E_0)+\mathbb{I}(Z=1)\kappa_v(v,E_v)+\mathbb{I}(Z=2)\kappa_u(u,E_u)$ where the latent variable $Z$ is independent of $(E_0,E_u,E_v)$ and satisfies that $\prob(Z=0)=\pi_0$, $\prob(Z=1)=\pi_v$, $\prob(Z=2)=\pi_u$. As such, Assumption \ref{asmp:transition}(i) and (ii) are automatically satisfied if $E_0, E_v$ and $E_u$ have sub-exponential tails, $\kappa_0$, $\kappa_u$ and $\kappa_v$ are Lipschitz continuous as functions of the error term and
\begin{eqnarray}\label{eqn:additivekappa}
    \sup_{a} \Mean \|\kappa_0(o,a,E_0)-\kappa_0(o',a,E_0)\|_2\le q\|o-o'\|_2,
\end{eqnarray}
for some $0\le q<1$. Notice that \eqref{eqn:additivekappa} is automatically satisfied for the auto-regressive model $O'=f(O,A)+g(E_0)$ for any $f$ such that $\sup_{a} |f(o,a)-f(o',a)|\le q\|o-o'\|_2$. Other examples are provided in \citet{diaconis1999iterated}. Assumption \ref{asmp:transition}(iii) requires the density of the latent factors and the initial observations to be upper bounded, thus yielding the uniform boundedness of marginal density functions of $\{O_{i,t}\}_{i,t}$. Assumption \ref{asmp:transition}(iv) lower bounds the second moment of the temporal difference error to invoke the martingale central limit theorem \citep{mcleish1974dependent} to establish the asymptotic normality of our estimator.

The first part of Assumption \ref{asmp:matrix} %implies that 
essentially requires $(NT)^{-1}\E(\bPhi_k^\top \bS_k\bPhi_k)$ %is bounded below and above by a constant, and thus allow us to show 
to be invertible. The second part is closely related to the irrepresentable or mutual incoherence condition in the variable selection literature for the selection consistency of the least absolute shrinkage and selection operator \citep{meinshausen2006high}. It imposes a norm constraint on the regression coefficients of the irrelevant predictors $\bm{\Phi}_{k-1}^{new}$ on the relevant predictors $\bm{\Phi}_k$. This type of assumption is necessary to ensure the consistency of the subsequent value estimator \citep{perdomo2022sharp}. Similar assumptions have been imposed in the statistics literature \citep{luckett2020estimating,shi2020statistical}. 

\textbf{\textit{Results}}. Finally, we present our theories. Recall that both $\eta_i^{\pi}$, $\eta_{i,t}^{\pi}$ as well as their estimators implicitly depend on the initial observation $O_{i,1}$. As such, it is proper to write them as functions of $O_{i,1}$, e.g., $\eta_{i,t}^{\pi}(o)=\Mean^{\pi}(R_{i,t}|O_{i,1}=o)$, $\widehat{\eta}_{i,t}^{\pi}(o)=\sum_a \widehat{Q}_{i,1,t}^{\pi}(o,a)\pi(a|o)$ ($\eta_{i}^{\pi}(o)$ and $\widehat{\eta}_i^{\pi}(o)$ can be similarly defined). For these values, instead of considering the differences $\widehat{\eta}_{i,t}^{\pi}(O_{i,1})-\eta_{i,t}^{\pi}(O_{i,1})$ and $\widehat{\eta}_{i,t}^{\pi}(O_{i,1})-\eta_{i}^{\pi}(O_{i,1})$, we focus on the aggregated differences $\int_{o\in \mathcal{O}}[\widehat{\eta}_{i,t}^{\pi}(o)-\eta_{i,t}^{\pi}(o)]do$ and $\int_{o\in \mathcal{O}}[\eta_{i}^{\pi}(o)-\widehat{\eta}_{i,t}^{\pi}(o)]do$ to eliminate the variability due to $O_{i,1}$. 

\begin{theorem}[Rates of Convergence]\label{thm:rateofconvergence}
{Assume Assumptions \ref{asmp: completeness}, \ref{asmp:basis}(i)--(iii), \ref{asmp:transition}(i)--(iii) and \ref{asmp:matrix} hold. Then with probability approaching $1$, we have for any $1 \leq i\leq N$ and $1 \leq t\leq T$,
\begin{align*}
    %\norm{\widehat\eta_{i,t}^{\pi}-
    %\eta_{i,t}^{\pi}}=O_p(N^{-1/2})+O_p(T^{-1/2}), \\
    &\max_{i,t} \left|\int_{o\in \mathcal{O}} [\widehat\eta_{i,t}^{\pi}(o)-\eta_{i,t}^{\pi}(o)]do\right|=O(L^{-p/d})+O(\sqrt{\log (N T)/N})+O(\sqrt{\log (NT)/T}),\\  
    %&\norm{\widehat\eta_{i}^{\pi}-\eta_{i}^{\pi}}=O_p(T^{-1/2}), \mbox{ } 
    &\max_{i} \left|\int_{o\in \mathcal{O}}[\widehat\eta_{i}^{\pi}(o)-\eta_{i}^{\pi}(o)]do\right|=O(L^{-p/d})+O(\sqrt{\log (NT)/T}),\\ 
    %&\norm{\widehat\eta_{t}^{\pi}-\eta_{t}^{\pi}}=O_p(N^{-1/2}), \mbox{ } 
    &\max_{t} |\widehat\eta_{t}^{\pi}-\eta_{t}^{\pi}|=O(L^{-p/d})+O(\sqrt{\log (NT)/N})\,\, \mbox{and }\,\, |\widehat\eta^{\pi}-\eta^{\pi}|=O(L^{-p/d})+O(\sqrt{\log (NT)/NT}).
\end{align*}
}
\end{theorem}

Theorem \ref{thm:rateofconvergence} highlights a noteworthy property of our method: the error bounds of value estimator depend solely on $p$, $d$, $L$, $N$ and $T$, and are independent of the number of backward inductions conducted. This is due to an important feature of our approach: the error term at the $k$th backward stage is of order $O(\pi_0^k)$ (as demonstrated by Lemma 1 in the supplementary article). Specifically, the error bounds for each value estimator comprise two components: the bias term $O(L^{-p/d})$ and the variance term $O(N^{-1/2}\sqrt{\log (NT)})$, $O(T^{-1/2}\sqrt{\log (NT)})$ or $O(\sqrt{\log (NT)/NT})$. The bias term quantifies the approximation error incurred by using linear sieves to approximate the underlying Q-function. Evidently, this bias term diminishes as the smoothness parameter $p$ increases. As such, it implies that the smoother the system dynamics are, the smaller the approximation error becomes. Moreover, for sufficiently large $L$, it is evident that due to aggregation over time and population, the average effect $\widehat{\eta}^{\pi}$ converges the fastest, whereas the individual- and time-specific effect $\widehat{\eta}^{\pi}_{i,t}$ demonstrates a relatively slower convergence. %Finally, the $\sqrt{\log(NT)}$ term arises from the application of Bernstein's concentration inequality. This is used to provide a simultaneous upper bound for all individual-specific effects and time-specific effects.

\begin{theorem}[Asymptotically Normality]\label{thm:asymnormal}{Assume Assumptions \ref{asmp: completeness} -- \ref{asmp:matrix} hold. Then when both $N$ and $T$ goes to infinity,
\begin{gather*}
\sqrt{\min (N,T)}\sigma_{\eta_{i,t}^{\pi}}^{-1}\int_{o\in \mathcal{O}}\left(\widehat\eta_{i,t}-\eta_{i,t}\right)do\xrightarrow{\enskip d\enskip} \mathcal{N}(0,1)\mbox{, } 
\sqrt{T}\sigma_{\eta_{i}^{\pi}}^{-1}\int_{o\in \mathcal{O}}(\widehat \eta_{i}^{\pi}-\eta_{i}^{\pi})do\xrightarrow{\enskip d \enskip} \mathcal{N}(0,1),\\
\sqrt{N}\sigma_{\eta_{t}^{\pi}}^{-1}(\widehat \eta_{t}^{\pi}-\eta_{t}^{\pi})\xrightarrow{\enskip d \enskip} \mathcal{N}(0,1)
\mbox{, and }\sqrt{NT}\sigma_{\eta^{\pi}}^{-1}(\widehat \eta^{\pi}-\eta^{\pi}) \xrightarrow{\enskip d\enskip} \mathcal{N}(0,1),
\end{gather*} where $\sigma_{\eta_{i,t}^{\pi}}, \sigma_{\eta_{i}^{\pi}}, \sigma_{\eta_{t}^{\pi}}$ and $\sigma_{\eta^{\pi}}$ are some quantities bounded from below and above (for a detailed formulation, refer to Section B.3 in the supplementary article). 
}
\end{theorem}

Theorem \ref{thm:asymnormal} establishes the asymptotic normality of the value estimators when both $N$ and $T$ diverges. It lays the foundations for statistical inference (e.g., constructing confidence intervals) of these policy values. Specifically, one could estimate the standard deviations $\sigma_{\eta_{i,t}^{\pi}}, \sigma_{\eta_{i}^{\pi}}, \sigma_{\eta_{t}^{\pi}}$, $\sigma_{\eta^{\pi}}$ from the data, and then employ these estimators to construct Wald-type confidence intervals. 
%As an alternative, bootstrap techniques or other resampling methods could potentially be leveraged for uncertainty quantification.

\section{Simulation Studies}\label{sec:sims}

In this section, we evaluate our proposed model-based and model-free approaches through extensive simulations. We begin by specifying the competing methods. Next, we
evaluate the performance of our model-free method using the RL benchmark dataset D4RL \citep{fu2020d4rl}. Finally, we investigate the sensitivity of the proposed model-free and model-based estimators to the additivity assumption. We focus on
evaluating the following four targets: (i) the average reward $\eta^{\pi}$; (ii) the $i$th subject’s average
reward aggregated over time $\eta^{\pi}_i$; (iii) the average reward in the population at time $t$ $\eta^{\pi}_t$; (iv) the $i$th subject’s expected reward at time $t$, denoted by $\eta^{\pi}_{i,t}$. Throughout, we evaluate these four targets for all subjects over the first five time periods in the offline dataset. Additional numerical results and details about the environments, can be found in Section C of the supplementary article.
%\res{We first demonstrate our methods under correct model specification \ref{eqn:model}, and explore three scenarios where the mixture probability $\pi_0$ varies.  To save space, we include the results of correctly specified models in Section \ref{sec: addsims} of the supplementary article. Then we conduct numerical studies to investigate the sensitivity of the proposed estimator to the additivity assumption in Section \ref{sec: sens}. Finally we evaluate the performance of our method using the benchmark dataset D4RL \citep{fu2020d4rl} at Section \ref{sec: D4RL}.}

\textit{\textbf{Competing methods}}. We compare our proposed approaches against the following methods, including two direct methods (DM), three importance sampling (IS) methods, three doubly robust (DR) methods, and one model-based (MB) method: \vspace{-0.5em}
\begin{enumerate}
    \item[(i)] DM1: an adaptation of fitted Q-evaluation \citep{le2019batch} to the average reward;\vspace{-0.5em}
    \item[(ii)] DM2 : an adaptation of Q-function based least-squares temporal difference \citep[see, e.g.,][]{shi2020statistical} to the average reward setting;\vspace{-0.5em}
    \item[(iii)] IS1: sequential IS that uses the product of IS ratios at each time to address the distributional shift between the behavior and target policies \citep{precup2000eligibility};\vspace{-0.5em}
    \item[(iv)] IS2: marginalized IS that replaces  the product of IS ratios with the marginalized IS ratio to break the curse of horizon \citep{liu2018breaking,kallus2020double};\vspace{-0.5em}
    \item[(v)] IS3: marginalized IS based on minimax weight learning \citep{uehara2020minimax};\vspace{-0.5em}
    \item[(vi)] DR1: a doubly robust method that employs the influence function developed by \citet{jiang2016doubly} to construct the estimator and uses approaches from DM1 and IS1 to compute the Q-function and the IS ratio; \vspace{-0.5em}
    \item[(vii)] DR2: a doubly robust method that employs the influence function developed by \citet{kallus2020double} to construct the estimator and uses approaches from DM1 and IS2 to compute the Q-function and the IS ratio; \vspace{-0.5em}
    \item[(viii)] DR3: a doubly robust method that employs the influence function developed by \citet{liao2022batch} to construct the estimator and uses approaches from DM2 and IS3 to compute the Q-function and the IS ratio; \vspace{-0.5em}
    \item[(ix)] MB: a standard model-based method developed in doubly homogeneous environments.\vspace{-0.5em} 
\end{enumerate}

We also remark that the three IS and DR methods primarily differ in their utilization of IS ratios. Specifically, IS1 and DR1 use the sequential IS ratio, requiring the environment to be individually homogeneous -- meaning all individuals' data trajectories share the same distribution. In contrast, IS2 and DR2 employ the marginalized IS ratio, whose validity additionally depends on the Markov assumption. Meanwhile, IS3 and DR3 apply another marginalized IS ratio that further requires the stationarity assumption. However, all the three aforementioned assumptions are violated in doubly inhomogeneous environments due to the presence of $\{U_i\}_i$ and $\{V_t\}_t$. Further details of these methods are relegated to Section C.1 of the supplementary article.

\textit{\textbf{Application to D4RL}}. 
D4RL consists of a collection of benchmark datasets specifically designed for evaluating RL algorithms. Its primary goal is to provide standardized and diverse data that assist researchers and practitioners in developing advanced methodologies for offline RL. We evaluate the performance of our method across four D4RL environments: Maze2D, Hopper, HalfCheetah, and Walker2d. For each environment, we further consider four distinct settings. For Maze2D, the settings differ in maze layouts and the level of difficulty in reaching the goal state. The four specific settings we consider include ``open", ``umaze", ``medium" and ``large". For HalfCheetah, Walker2D and Hopper, the settings are defined by varying behavior policies, and we consider the four settings labeled ``noisy", ``medium", ``medium-replay" and ``medium-expert". The datasets can be directly downloaded from \url{http://rail.eecs.berkeley.edu/datasets/offline_rl/}. More details can be found in Section C.2 of the supplementary article and the D4RL Wiki page\footnote{\url{https://github.com/Farama-Foundation/d4rl/wiki/Tasks}.}.

In all settings, the target policy we aim to evaluate is fixed to a randomized policy that follows a uniform distribution across the action space. To simulate doubly inhomogeneous environments, we inject two-way fixed effects into the original reward $R_{i,t}$ from the D4RL datasets, leading to the modified reward $\widetilde{R}_{i,t} = R_{i,t} + \cos(t) + \sin(i)$. All observations and actions from the original data remain unchanged. To ensure fair comparison, the Q-functions in the proposed, direct and doubly robust methods are all modeled via linear sieves with a quadratic basis function. For IS and DR, we use a conditional Gaussian model with a linear conditional mean function and constant variance to approximate the behavior policy.

\spacingset{1.1}
\begin{table}[t] 
\caption{MSEs of the estimated value (four targets) using our proposed methods and other competing methods for Maze2D and Halfcheetah with $N=T=20$ over 20 replications. The best method with smallest MSE in each column were highlighted with blue. } \label{table:maze1}
\centering
\begin{threeparttable}
\resizebox{\textwidth}{!}{%
\begin{tabular}[t]{lrrrrrrrrrrrrrrrr} 
\toprule
\multicolumn{1}{c}{} & \multicolumn{4}{c}{Maze2D-open} & \multicolumn{4}{c}{Maze2D-umaze} & \multicolumn{4}{c}{Maze2D-medium} & \multicolumn{4}{c}{Maze2D-large}\\
\cmidrule(l{4pt}r{4pt}){2-5} \cmidrule(l{4pt}r{4pt}){6-9} \cmidrule(l{4pt}r{4pt}){10-13} \cmidrule(l{4pt}r{4pt}){14-17}
& $\eta^\pi$ & $\eta^\pi_{i}$  & $\eta^\pi_{t}$ & $\eta^\pi_{i,t}$ & $\eta^\pi$ & $\eta^\pi_{i}$  & $\eta^\pi_{t}$ & $\eta^\pi_{i,t}$ &$\eta^\pi$ & $\eta^\pi_{i}$  & $\eta^\pi_{t}$ & $\eta^\pi_{i,t}$ &$\eta^\pi$ & $\eta^\pi_{i}$  & $\eta^\pi_{t}$ & $\eta^\pi_{i,t}$ \\
\midrule
P1  &\cellcolor{lightblue} 0.01 &\cellcolor{lightblue}0.45 &\cellcolor{lightblue} 0.30 &\cellcolor{lightblue}0.75 &\cellcolor{lightblue}0.02& \cellcolor{lightblue}0.47& \cellcolor{lightblue}0.28 &\cellcolor{lightblue}0.73& \cellcolor{lightblue} 0.00 &\cellcolor{lightblue}0.45& \cellcolor{lightblue}0.31& \cellcolor{lightblue}0.76 &\cellcolor{lightblue}0.00 &\cellcolor{lightblue}0.45 &0.32 &\cellcolor{lightblue}0.77 \\

DM1 & 0.01& 0.49 &0.34& 0.83&
0.03 &0.52 &0.33 &0.82&
 0.01& 0.51& 0.35 &0.85&
 0.00& 0.50 &0.36 &0.85\\

DM2 & 3.75& 4.25& 3.72& 4.23&
 2.98 &3.49& 3.93 &4.43&
 0.75& 1.26& 1.12 &1.64&
 0.55& 1.07& 0.96& 1.48\\

IS1 &0.66 &1.17& 1.26 &3.63&
 0.42 &0.93& 0.39 &2.06&
 0.35& 0.87& 0.62 &2.56&
 0.62 &1.13& 1.12 &3.34\\

IS2 &1.52 & 2.03&  6.10 &10.12&
 1.81 &2.32& 4.65& 8.06&
 0.93 &1.44& 3.43& 6.67&
 1.28& 1.80 &5.22& 8.94\\

IS3 & 0.01& 0.52& 0.35 &0.85&
 0.03 &0.54 &0.33 &0.84&
 0.01& 0.52& 0.35 &0.87&
 0.00& 0.52 &0.36& 0.87\\

DR1  &0.25 &2.99& 0.44& 7.03&
 0.99 &12.81 & 3.11& 60.60&
 0.15 &1.80 &0.38& 7.45&
 0.21& 1.41& \cellcolor{lightblue}0.28 &4.31\\

DR2 & 0.25 & 3.09 & 1.16& 13.04&
 0.13 &2.68 &0.65& 9.86&
 0.18& 2.35 &0.64 &8.82&
 0.21 &1.95& 0.64& 8.06\\

DR3  &0.01 &0.51& 0.36& 0.86&
 0.03& 0.54& 0.33 &0.84&
 0.01& 0.52& 0.36& 0.87&
 0.00 &0.52& 0.36& 0.88\\
\midrule
\multicolumn{1}{c}{} & \multicolumn{4}{c}{Halfcheetah-medium} & \multicolumn{4}{c}{Halfcheetah-noisy} & \multicolumn{4}{c}{Halfcheetah-medium-replay} & \multicolumn{4}{c}{Halfcheetah-medium-expert}\\
\cmidrule(l{4pt}r{4pt}){2-5} \cmidrule(l{4pt}r{4pt}){6-9} \cmidrule(l{4pt}r{4pt}){10-13} \cmidrule(l{4pt}r{4pt}){14-17}
& $\eta^\pi$ & $\eta^\pi_{i}$  & $\eta^\pi_{t}$ & $\eta^\pi_{i,t}$ & $\eta^\pi$ & $\eta^\pi_{i}$  & $\eta^\pi_{t}$ & $\eta^\pi_{i,t}$ & $\eta^\pi$ & $\eta^\pi_{i}$  & $\eta^\pi_{t}$ & $\eta^\pi_{i,t}$ & $\eta^\pi$ & $\eta^\pi_{i}$  & $\eta^\pi_{t}$ & $\eta^\pi_{i,t}$ \\
\midrule
P1  & 0.06 & \cellcolor{lightblue}0.56 & \cellcolor{lightblue} 0.44 & \cellcolor{lightblue}0.95
&\cellcolor{lightblue} 0.06 & \cellcolor{lightblue}0.57 & \cellcolor{lightblue} 0.44 &\cellcolor{lightblue} 0.95
& \cellcolor{lightblue}0.05 & \cellcolor{lightblue}0.57 & \cellcolor{lightblue}0.46 & \cellcolor{lightblue}0.97
& 0.06 & \cellcolor{lightblue}0.57 & \cellcolor{lightblue}0.45 & \cellcolor{lightblue}0.96 \\

DM1 & 0.06 & 0.56 & 0.44 & 0.95
& 0.06 & 0.57 & 0.45 & 0.96
& 0.05 & 0.57 & 0.46 & 0.97
& 0.06 & 0.57 & 0.45 & 0.96 \\

DM2 & 1.46 & 1.96 & 1.92 & 2.43
&0.36& 0.87& 0.87 &1.38
& 0.50 & 1.02 & 1.01 & 1.52
& 1.46 & 1.97 & 1.92 & 2.44 \\

IS1 & 0.06 & 0.57 & 0.47 & 0.99
& 0.27 & 0.77 & 1.07 & 1.89
& 0.08 & 0.60 & 0.46 & 1.06
& 0.06 & 0.58 & 0.48 & 1.00 \\

IS2 & \cellcolor{lightblue}0.05 & 0.56 & 0.45 & 0.96
&  0.63 &1.14& 2.69 &3.46
& 0.10 & 0.62 & 0.95 & 1.87
& \cellcolor{lightblue}0.05 & 0.57 & 0.46 & 0.97 \\

IS3 & 0.05 & 0.56 & 0.46 & 0.97
& 0.06 &0.57 &0.48& 0.99
& 0.05 & 0.57 & 0.46 & 0.97
& 0.06 & 0.57 & 0.46 & 0.98 \\

DR1  & 0.15 & 0.77 & 0.86 & 1.92
& 3.82 &3.13& 4.25& 6.34
& 0.38 & 2.88 & 1.77 & 12.80
& 0.15 & 0.78 & 0.87 & 1.92 \\

DR2 & 0.06 & 0.57 & 0.44 & 0.96
& 0.06 &0.60 &0.45 &1.08
& 0.10 & 0.80 & 0.52 & 2.50
& 0.06 & 0.57 & 0.45 & 0.97 \\

DR3  & 0.08 & 0.59 & 0.51 & 1.02
& 0.22 &0.73& 0.70 &1.21
& 0.05 & 0.57 & 0.46 & 0.98
& 0.09 & 0.60 & 0.51 & 1.02 \\
\bottomrule
\end{tabular}
}
\end{threeparttable}
\end{table}

\begin{table}[t] 
\caption{MSEs of the estimated value (four targets) using our proposed methods with other competing methods for Walker2D and Hopper with $N=T=20$ over $20$ replications. The best method with the smallest MSE in each column is highlighted in blue. }\label{table:walker1}
\centering
\begin{threeparttable}
\resizebox{\textwidth}{!}{%
\begin{tabular}[t]{lrrrrrrrrrrrrrrrr} 
\toprule
\multicolumn{1}{c}{} & \multicolumn{4}{c}{Walker2D-medium} & \multicolumn{4}{c}{Walker2D-noisy} & \multicolumn{4}{c}{Walker2D-medium-replay} & \multicolumn{4}{c}{Walker2D-medium-expert}\\
\cmidrule(l{4pt}r{4pt}){2-5} \cmidrule(l{4pt}r{4pt}){6-9} \cmidrule(l{4pt}r{4pt}){10-13} \cmidrule(l{4pt}r{4pt}){14-17}
& $\eta^\pi$ & $\eta^\pi_{i}$  & $\eta^\pi_{t}$ & $\eta^\pi_{i,t}$ & $\eta^\pi$ & $\eta^\pi_{i}$  & $\eta^\pi_{t}$ & $\eta^\pi_{i,t}$ & $\eta^\pi$ & $\eta^\pi_{i}$  & $\eta^\pi_{t}$ & $\eta^\pi_{i,t}$ & $\eta^\pi$ & $\eta^\pi_{i}$  & $\eta^\pi_{t}$ & $\eta^\pi_{i,t}$ \\
\midrule
P1  & \cellcolor{lightblue}0.51 & \cellcolor{lightblue} 1.03 & \cellcolor{lightblue}0.55 & \cellcolor{lightblue} 1.07
& 0.52 & 1.03 & 0.55 & \cellcolor{lightblue}1.07
& 0.53 & \cellcolor{lightblue}1.03 & \cellcolor{lightblue}0.55 & \cellcolor{lightblue} 1.06
& \cellcolor{lightblue}0.52 & \cellcolor{lightblue}1.03 & \cellcolor{lightblue}0.55 & \cellcolor{lightblue}1.07 \\

DM1 & 0.52 & 1.03 & 0.56 & 1.08
& 0.52 & 1.04 & 0.55 & 1.07
& 0.53 & 1.04 & 0.56 & 1.07
& 0.52 & 1.04 & 0.56 & 1.08 \\

DM2 & 38.93 & 39.45 & 38.60 & 39.12
& 22.15 &22.66& 23.29 &23.80
& 31.70 & 32.21 & 39.07 & 39.58
& 38.94 & 39.46 & 38.60 & 39.11 \\

IS1 & 0.52 & 1.03 & 0.57 & 1.09
& 8.31 & 8.82 &51.74 &57.08
& 0.53 & 1.04 & 0.56 & 1.07
& 0.53 & 1.04 & 0.57 & 1.08 \\

IS2 & 0.52 & 1.03 & 0.57 & 1.09
& \cellcolor{lightblue}0.36 &\cellcolor{lightblue}0.87& \cellcolor{lightblue}0.40 &1.42
& 0.53 & 1.04 & 0.55 & 1.08
& 0.53 & 1.04 & 0.57 & 1.08 \\

IS3 & 0.52 & 1.03 & 0.57 & 1.09
& 0.53 & 1.04 & 0.57 & 1.08
& 0.53 & 1.04 & 0.57 & 1.07
& 0.53 & 1.04 & 0.57 & 1.08 \\

DR1 & 0.56 & 1.07 & 1.02 & 1.55
& 3.09  & 6.55 & 22.45 & 28.16
& 0.59 & 1.12 & 0.98 & 1.58
& 0.56 & 1.08 & 1.02 & 1.55 \\

DR2 & 0.52 & 1.03 & 0.56 & 1.08
& 0.48 &1.00 &0.48& 1.53
& 0.53 & 1.06 & 0.56 & 1.18
& 0.52 & 1.04 & 0.56 & 1.08 \\

DR3 & 0.52 & 1.03 & 0.56 & 1.09
& 0.53& 1.04 &0.57 &1.08
& \cellcolor{lightblue}0.52 & 1.03 & 0.56 & 1.07
& 0.52 & 1.04 & 0.56 & 1.08 \\

\midrule 
\multicolumn{1}{c}{} & \multicolumn{4}{c}{Hopper-medium} & \multicolumn{4}{c}{Hopper-noisy} & \multicolumn{4}{c}{Hopper-medium-replay} & \multicolumn{4}{c}{Hopper-medium-expert}\\
\cmidrule(l{4pt}r{4pt}){2-5} \cmidrule(l{4pt}r{4pt}){6-9} \cmidrule(l{4pt}r{4pt}){10-13} \cmidrule(l{4pt}r{4pt}){14-17}
& $\eta^\pi$ & $\eta^\pi_{i}$ & $\eta^\pi_{t}$ & $\eta^\pi_{i,t}$ & $\eta^\pi$ & $\eta^\pi_{i}$ & $\eta^\pi_{t}$ & $\eta^\pi_{i,t}$ & $\eta^\pi$ & $\eta^\pi_{i}$ & $\eta^\pi_{t}$ & $\eta^\pi_{i,t}$ & $\eta^\pi$ & $\eta^\pi_{i}$ & $\eta^\pi_{t}$ & $\eta^\pi_{i,t}$ \\
\midrule
P1 & \cellcolor{lightblue} 0.43 & \cellcolor{lightblue} 0.92 & \cellcolor{lightblue}0.76 & \cellcolor{lightblue}1.25 & 0.38& 0.87& 0.67 &1.17 & 0.43 & 0.92 & 0.76 & \cellcolor{lightblue}1.25 & 0.42 & \cellcolor{lightblue}0.90 & \cellcolor{lightblue}0.76 &\cellcolor{lightblue} 1.25 \\

DM1 & 0.44 & 0.96 & 0.79 & 1.30 & 0.39 &0.90 &0.69 &1.21 & 0.44 & 0.96 & 0.80 & 1.31 & 0.44 & 0.94 & 0.80 & 1.31 \\

DM2 & 40.91 & 41.44 & 40.98 & 41.50 & 4.02 &4.53& 4.29& 4.81& 26.61 & 27.13 & 27.09 & 27.60 & 21.99 & 22.49 & 22.29 & 22.80 \\

IS1 & 0.48& 1.00 &0.86 &1.39&
 0.48 &1.00 &2.12 &2.85&
 \cellcolor{lightblue} 0.37& \cellcolor{lightblue}0.89& \cellcolor{lightblue}0.70 &1.42&
 0.55& 1.05& 1.12& 1.93 \\

IS2 & 0.48 &1.00& 0.87& 1.39&
 0.42 &0.93& 1.17 &1.88&
 0.41 &0.93& 0.78 &1.58&
 0.46& 0.96 &0.82& 1.61 \\

IS3 & 0.46 & 0.98 & 0.86 & 1.37 
& 0.44 &0.95& 0.83 &1.35
& 0.46 & 0.98 & 0.86 & 1.37 & 0.45 & 0.96 & 0.85 & 1.36 \\

DR1 & 0.45 & 0.97 & 0.81 & 1.32 
& 0.41 &1.23& 0.75 &2.54 & 
0.50 & 1.11 & 0.82 & 1.91 & \cellcolor{lightblue}0.39 & 1.47 & 0.81 & 4.77 \\

DR2 & 0.44 & 0.95 & 0.79 & 1.30 
& 0.59 &1.39 &1.05 &3.18
& 0.46 & 1.04 & 0.81 & 2.33 
& 0.51 & 1.18 & 0.84 & 2.43 \\

DR3 & 3.48 & 4.00 & 3.88 & 4.39 & \cellcolor{lightblue} 0.22 &\cellcolor{lightblue} 0.74 & \cellcolor{lightblue}0.60 & \cellcolor{lightblue}1.12 & 0.93 & 1.45 & 1.30 & 1.81 & 8.42 & 8.92 & 8.75 & 9.26 \\
\bottomrule
\end{tabular}
}
\end{threeparttable}
\end{table}

\spacingset{1.7}
MSEs of various model-free estimators are reported in Tables \ref{table:maze1} and \ref{table:walker1}\footnote{To enhance clarity, we remove the two model-based methods in the tables, as their performance are much worse than the model-free methods, due to the difficulty in accurately modeling the complex transition function in D4RL.}. Recall that for each environment, we consider four settings, each containing four evaluation targets. This yields 16 cases for each environment. 
Overall, the proposed model-free method (denoted by P1) %outperforms the competing methods 
achieves the best performance in most cases: 
\begin{itemize}
    \item In Maze2D, our proposed method ranks first in 15 out of 16 cases;
    \item In HalfCheetah, our method ranks first in 14 out of 16 cases;
    \item In Walker2D, our proposed method ranks first in 12 out of 16 cases;
    
    \item In Hopper, our method ranks first in 8 out of 16 cases.
\end{itemize}
 %in hopper, our proposed method ranks first in 8 out of 12 estimands (most among all the methods); in walker2D, our method ranks first in 11 out of 12 estimands (most among all the methods); while in halfCheetah, our method ranks first in 10 out of 12 estimands (most among all the methods). 
 Meanwhile, there are a few exceptions: %. For example, 
 In the Hopper-noisy setting, DR3 outperforms our method for evaluating all the four estimands. Likewise, for Hopper-medium-replay, IS1 achieves the best performance for estimating $\eta^\pi$, $\eta^\pi_{i}$ and $\eta^\pi_{t}$. Despite these specific cases, our method %generally 
 exhibits superior performance in general. %across a broader range of settings compared to the alternatives. 
It is also worth noting that our proposed method is not overly sensitive to diverse behavior policies. In contrast, competing methods like IS1 and IS2, which require to learn the behavior policy, vary considerably across different settings. 

\textbf{\textit{Sensitivity analysis}}. We have designed four synthetic environments with binary actions and continuous observations to investigate the sensitivity of our model-free (P1) and model-based (P2) estimators to the additivity assumption. These environments differ from D4RL in that, unlike the D4RL where only the reward is doubly inhomogeneous, in these environments, we introduce latent factors into the transition function to make it doubly inhomogeneous as well.

Specifically, we consider two reward models: an additive model \eqref{eqn:reward} and a factor model \eqref{eqn:rewardfactor}. In the additive model, the two-way fixed effects $\theta_i$ and $\lambda_t$ are set to $\sin(i)$ and $\cos(t)$, respectively. In the factor model, we set $\bgamma_i=\left(\sin(i),\sin(2i),\sin(3i)\right)^\top$ and $\balpha_t=(\cos(t),\cos(2t),\cos(3t))^\top$. In both models, the reward function is fixed to $r_1(o,a)=a-0.25o$ and residuals $\varepsilon_{i,t}$s are i.i.d. Gaussian random errors with mean zero and variance $0.25$.

% . Additionally, , respectively. In the interactive model, the 
Moreover, we consider three transition models: (i) an additive model where 
\spacingset{1.3}
$O_{i,t+1}=-0.25O_{i,t}+A_{i,t}+\sin(i)+\cos(t)+e_{i,t}$; (ii) a factor model where $O_{i,t+1}=-0.25O_{i,t}+A_{i,t}+\bgamma_i^\top \balpha_t+e_{i,t}$; (iii) a regime switching model where\vspace{-0.5em}
\begin{gather*}
O_{i,t+1}= \left\{\begin{array}{ll}\vspace{-0.2em}
	-0.25O_{i,t}+A_{i,t}+2e_{i,t}, & \text{if both $i$ and $t$ is odd},\\ \vspace{-0.2em}
   O_{i,t}-A_{i,t}+e_{i,t}, & \text{if $i$ is odd, and $t$ is even},\\\vspace{-0.2em}
   0.25O_{i,t}-A_{i,t}+2e_{i,t}, & \text{if $i$ is even, and $t$ is odd},\\\vspace{-0.2em}
    -O_{i,t}+A_{i,t}+e_{i,t}, & \text{otherwise}.
		 \end{array} \;\;\right.
\end{gather*}\spacingset{1.7}
Similarly, $e_{i,t}$s are set to i.i.d. Gaussian errors with mean zero and variance $0.25$.

\spacingset{1.1}
\begin{table}[t]
\caption{ A summary of environments in the sensitivity analysis. } \label{table:sensi}
\centering
\footnotesize
\begin{tabular}[t]{lrrrr}
\toprule
Environment & I & II & III & IV  \\
\midrule
Reward & additive & additive & factor & factor  \\
Transition & regime switching & factor & additive & factor  \\
\bottomrule
\end{tabular}
\end{table}

\spacingset{1.1}
\begin{table}[t] 
\caption{MSEs of the estimated value (four targets) using our proposed methods with other competing methods. The best method with the smallest MSE in each column is highlighted in blue. P1 and P2 are our proposed model-free and model-based methods, respectively.} 
\label{table:sens}
\centering
\begin{threeparttable}
\resizebox{0.9\textwidth}{!}{%
\begin{tabular}[t]{lrrrrrrrrrrrrrrrr} 
\toprule
\multicolumn{1}{c}{} & \multicolumn{4}{c}{Scenario 1} & \multicolumn{4}{c}{Scenario 2} & \multicolumn{4}{c}{Scenario 3} & \multicolumn{4}{c}{Scenario 4} \\
\cmidrule(l{4pt}r{4pt}){2-5} \cmidrule(l{4pt}r{4pt}){6-9} \cmidrule(l{4pt}r{4pt}){10-13} \cmidrule(l{4pt}r{4pt}){14-17} 
& $\eta^\pi$ & $\eta^\pi_{i}$ & $\eta^\pi_{t}$ & $\eta^\pi_{i,t}$ & $\eta^\pi$ & $\eta^\pi_{i}$ & $\eta^\pi_{t}$ & $\eta^\pi_{i,t}$ & $\eta^\pi$ & $\eta^\pi_{i}$ & $\eta^\pi_{t}$ & $\eta^\pi_{i,t}$ & $\eta^\pi$ & $\eta^\pi_{i}$ & $\eta^\pi_{t}$ & $\eta^\pi_{i,t}$  \\
\midrule
P1 & \cellcolor{lightblue}0.01 & \cellcolor{lightblue}0.48 & 0.17 & 3.54
   & 0.66 & \cellcolor{lightblue}0.73 & \cellcolor{lightblue}0.10 & \cellcolor{lightblue}1.78
   & 0.41 & \cellcolor{lightblue}0.51 & \cellcolor{lightblue}0.07 & 4.65
   & 0.03 &0.17& \cellcolor{lightblue}0.05& 9.00 \\
P2 & 0.04 & 0.76 & \cellcolor{lightblue}0.10 & \cellcolor{lightblue} 3.43
   & 0.33 & 1.04 & 0.35 & 2.04
   & 0.47 & 0.60 & 0.10 & 4.70
   &  0.09& 0.28 &0.11 &9.11\\
MB & 0.01 & 1.56 & 0.84 & 4.80
   & 0.14 & 1.72 & 2.17 & 4.52
   & 0.09 & 0.64 & 0.31 & 4.82
   & \cellcolor{lightblue}0.01 &0.15& 0.11 &9.12 \\
DM1 & 0.01 & 1.40 & 0.85 & 4.02
    & \cellcolor{lightblue}0.01 & 1.26 & 1.26 & 3.06
    &\cellcolor{lightblue} 0.04 & 0.51 & 0.37 & 4.36
    & 0.02 &\cellcolor{lightblue}0.02 &0.08 &8.37 \\
DM2 & 0.37 & 1.77 & 0.98 & 4.16
    & 0.39 & 1.64 & 0.85 & 2.31
    & 0.81 & 1.27 & 0.51 & \cellcolor{lightblue}4.08
    & 0.77& 0.77 &0.73 &8.25 \\
IS1 & 0.20 & 1.60 & 0.58 & 5.25
    & 0.84 & 2.08 & 0.55 & 3.41
    & 0.63 & 1.09 & 0.63 & 4.74
    & 0.30& 0.30 &0.78 &8.83 \\
IS2 & 0.05 & 1.45 & 0.13 & 4.82
    & 1.26 & 2.50 & 0.31 & 3.43
    & 0.93 & 1.40 & 0.33 & 4.48
    & 0.41 &0.41& 0.36 &\cellcolor{lightblue}8.23 \\
IS3 & 2.89 & 4.28 & 3.14 & 6.32
    & 7.04 & 8.29 & 4.63 & 6.09
    & 7.47 & 7.94 & 5.17 & 8.74
    &  6.48 & 6.49 & 5.72& 13.24 \\
DR1 & 0.16 & 1.56 & 0.35 & 5.14
    & 0.53 & 1.78 & 0.29 & 3.00
    & 0.67 & 1.14 & 0.33 & 4.54
    & 0.24 &0.24 &0.37 &8.36 \\
DR2 & 0.23 & 1.63 & 0.85 & 6.32
    & 0.58 & 1.82 & 0.25 & 3.67
    & 0.95 & 1.41 & 0.60 & 5.07
    & 0.22& 0.22& 0.37 &8.38 \\
DR3 & 2.21 & 3.61 & 2.54 & 5.72
    & 7.02 & 8.26 & 4.61 & 6.08
    & 6.84 & 7.31 & 4.66 & 8.24
    &  5.54 & 5.54 & 4.86 &12.38\\
\bottomrule
\end{tabular}
}
\end{threeparttable}
\end{table}

\spacingset{1.7}
Table \ref{table:sensi} summarizes reward and transition models of the four environments. It can be seen that in each environment, either the reward or the transition model does not satisfy the additive structure, leading to the violation of the proposed model assumption. For all the environments% outlined above
, we set $N=40$, $T=40$. %and $r_1(o,a)=2o+3a$. For the scenarios including the factor structure, we set $h=3$, $\bgamma_i=\left(\sin(i),\sin(2i),\sin(3i)\right)^\top$, and $\balpha_t=(\cos(t),\cos(2t),\cos(3t))^\top$. 
The behavior policy is a uniform random policy whereas the target policy %we aim to evaluate 
is %also 
another random policy where $\pi(1|o)=0.8$ for any $o$. The results are reported in Table \ref{table:sens}. In first three environments %show that even 
where the additivity assumption holds for either the reward or the transition model, our proposed method generally outperforms the competing methods in estimating $\eta^{\pi}_i$, $\eta^{\pi}_t$ and $\eta^{\pi}_{i,t}$. 
In the last environment where both models are interactive, our proposed method no longer dominates other methods, but its performance remains comparable to those of the best competing methods. 
%that of the standard model-based methods and DM1, and it still outperforms all other methods.

\section{Real Data Analysis}\label{sec:real}

In this section, we apply our proposed method to a sepsis dataset from MIMIC-III \citep{johnson2016mimic}, a database that contains information on critical care patients from Beth Israel Deaconess Medical Center in Boston, Massachusetts. As mentioned earlier, the heterogeneity in patients' response to treatment \citep{evans2021surviving}, along with potentially non-stationary environments makes it difficult to consistently assess the impact of conducting a given target policy on patient outcomes. 

We focus on a subset of patients who received treatments sequentially over 20 stages. The primary outcome in this analysis is the sequential organ failure assessment (SOFA) score \citep{jones2009sequential}, which monitors the progression of organ failure over time and measures the degree of organ dysfunction or failure in critically ill patients. A higher SOFA score indicates a higher risk of mortality. At any time point $t$, we consider a binary treatment $A_t \in \{0,1\}$ where $A_t=1$ indicates that the patient received an intravenous fluid intervention with a dose greater than the median value for the group of patients being studied, and $A_t=0$ otherwise. In previous studies, \citet{zhou2022optimizing} examined joint action spaces with both vasopressors and intravenous fluid interventions. %However, in this study, 
We focus solely on the intravenous fluid intervention in light of the findings of \citet{zhou2022optimizing}, which detected a limited impact of vasopressors.

\begin{figure}[t]
    \centering
    \spacingset{1.2}
    \includegraphics[height=4cm, width=0.5\textwidth]{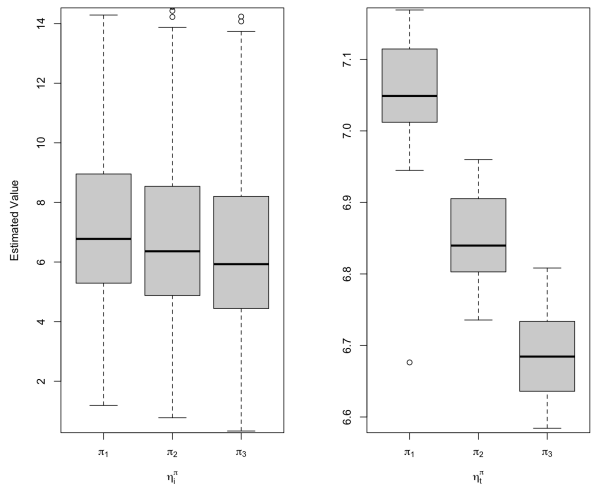}
    \caption{The estimated value function of $\eta^{\pi}_i$ and $\eta^{\pi}_t$ under three target policies, where $\pi_1$ is the always high dose policy, $\pi_2$ is always low dose policy, and $\pi_3$ is the tailored policy.}
    \label{fig:mimic}
\end{figure}

The following five covariates are included in the analysis: gender, age, the Elixhauser comorbidity index, weight, and the systemic inflammatory response syndrome score. Three deterministic policies were evaluated using our proposed methods: (i) always administering a high dose, (ii) always administering a low dose, and (iii) administering a low dose when the SOFA score is less than 11, and a high dose otherwise. The third policy is tailored to the SOFA score, taking into account evidence that a SOFA score of more than 11 is associated with a $100\%$ mortality rate \citep{jones2009sequential}. To estimate the Q-function, we employed a second-order degree polynomial two-way fixed effects model at each iteration. The average value estimators for the three policies are as follows: 7.26 (always high dose), 6.85 (always low dose), and 6.51 (tailored by SOFA score). These results indicate that the tailored policy is the most effective policy as it yields the lowest estimated SOFA score. Figure \ref{fig:mimic} summarizes the estimated $\eta^{\pi}_i$s and $\eta^{\pi}_t$s, clearly demonstrating that the tailored policy outperforms the other two policies, while the always high dose policy performs the worst. Our conclusion is in line with these existing results, which recommend the low dose policy over the high dose policy. It is also consistent with physicians’ recommendations in the behavior data \citep{zhou2022optimizing}.

\section*{Acknowledgements}
Shi's research is partially supported by an EPSRC grant EP/W014971/1. The authors thank the editor, AE, and the reviewers for their constructive comments, which have led to a significant improvement of the earlier version of this article.

\section*{Disclosure Statement}
The authors report there are no competing interests to declare.

\bibliographystyle{apalike}

\bibliography{references}

\end{document}